\documentclass[structabstract]{aa}  
\usepackage{graphicx}
\usepackage{txfonts}
\usepackage{multirow}
\usepackage{enumerate}
\usepackage{natbib} 
\begin{document}
   \title{X-ray variability of pre-main-sequence stars}

   \subtitle{Toward an explanation of the different X-ray properties of classical and weak-line T Tauri stars}

\author{E. Flaccomio\inst{1} \and G. Micela\inst{1} \and S. Sciortino\inst{1}}
\institute{INAF - Osservatorio Astronomico di Palermo, 
  Piazza del Parlamento 1, I-90134 Palermo, Italy \\ \email{E. Flaccomio, ettoref@astropa.unipa.it}
}  

   \date{Received Month XX, YYYY; accepted Month ZZ, HHHH}

 
  \abstract
   {The first million years of the pre-main-sequence (PMS) evolution of low-mass stars are characterized by magnetospheric accretion, a circumstellar proto-planetary disk, and the processes leading to its dispersal. Among these, photo-evaporation caused by strong X-ray emission from the central star is probably significant. Several aspects of the X-ray emission from coronae and accretion shocks remain mysterious, e.g.,  whether and how much accretion affects coronal emission.}
   {We studied the X-ray variability of  $\sim$1Myr old low-mass PMS stars as a function of timescale, stellar rotation, and stellar characteristics to gain insights into the working of PMS coronae, their X-ray emission, and the circumstellar environment in which they are immersed.}
   {We have exploited the $\sim$850\,ksec long {{\em Chandra}} observation of the Orion Nebula Cluster obtained by the COUP collaboration in January 2003,  and statistically analyzed the X-ray lightcurves of low-mass stars in several subsamples.  Our main focus was to characterize the different X-ray behavior of stars with and without circumstellar accretion disks, and to infer the physical mechanism responsible for the observed variability.}
   {Accreting stars (classical T Tauri stars, CTTSs) are found to be more variable than non-accreting stars (weak-line T Tauri stars, WTTSs) at all timescales and in all X-ray energy bands considered. Variability is seen to increase with timescale, up to the longest probed, $\sim$10 days. Signatures of rotational modulation are observed for both CTTSs and WTTSs, and most clearly for CTTSs in the soft X-ray band. Lower mass stars are more variable than higher mass stars.}
   {We propose that the difference in variability between CTTSs and WTTSs may be explained assuming that the X-ray emission of CTTS is affected by time-variable absorption due to circumstellar structures, such as warps in the inner disk and/or accretion streams. This suggestion is appealing because,  in the hypothesis that the coronae of CTTSs and WTTSs are similar, it may also explain why CTTSs have lower and more scattered X-ray emission levels compared to WTTSs.}


 \maketitle
 
%

\section{Introduction}

The evolution of young stars in the first few million years after they
emerge from their protostellar dusty envelopes is characterized by the
presence of circumstellar disks and intense magnetic fields. These
latter crucially regulate the interaction between the star and the disk,
they mediate for instance the angular momentum transfer between the two and
channel streams of accreting material from the disk onto the stellar
surface \citep[e.g.][]{bou07a}.  This material is shocked in the impact,
producing optical, UV, and soft X-ray ``excesses'' \citep{cal98}.
Magnetic fields are also responsible for the confinement and heating of
X-ray-bright coronae \citep{fei99}, whose high energy radiation, along
with that from accretion shocks, is likely to significantly influence
the evolution of circumstellar disks: soft X-rays ionize and heat parts
of the disk and may be the main cause of disk photo-evaporation and,
consequently, of their eventual dispersal \citep[e.g.][]{gla04,owe11}.

The above processes are highly dynamical. Strong variability at all
wavelengths and on many different timescales is indeed a distinctive
trait of young accreting pre-main-sequence (PMS) stars
\citep{joy45,ale10}, also known as classical T Tauri stars (CTTS).
Weak-line T Tauri stars (WTTS), similarly young stars for which
accretion signatures are not observed, represent almost certainly the
following stage of PMS evolution, once the circumstellar accretion has
stopped and the dusty disk has dissipated or is quickly dissipating
\citep{fei81, lad87}. Compared to CTTSs, WTTSs are much much less
variable in the optical band and their lightcurves show mostly periodic
variations attributable to the modulation of cool stellar spots
\citep{ale10}.

The mechanism responsible for the strong and irregular optical
variability of CTTSs, although likely related to accretion, remains
elusive. The two main hypotheses are $i$) variability of the emission
from the accretion shock(s), both intrinsic and caused by rotational
modulation, or $ii$) variable absorption of the photosphere, or part of
the photosphere, from unstable and optically thick accretion streams
and/or warps in the circumstellar disk. \citet{gra07,gra08}, for
example, suggested from a statistical study of CTTS that in about 25\%
of the cases variable absorption is the likely cause of optical
variability, while time-variable accretion is favored in the remaining
cases. However, owing to the similar effects of spots and absorption on
the broad-band lightcurves, the absorption scenario could not be
excluded in most cases. The thorough study of the CTTS \object{AA\,Tau}
by Bouvier and collaborators \citep{bou99,bou03,men03,bou07,gro07} has
shown that for this near edge-on star-disk system, the strong, irregular
optical variability is readily explained by occultation of a significant
fraction of the stellar surface by a warp in the inner disk. This warp,
located at the corotation radius, produces quasi-periodic eclipses with
the period of the stellar rotation. \citet{ale10}, using high-quality
optical lightcurves of stars in the NGC\,2264 region obtained with the
CoRoT satellite, have recently proposed that an \object{AA\,Tau}-like
scenario might indeed explain the optical variability of at least
$\sim$30\% of the CTTS. The AA Tau-like phenomenology  was also
confirmed by \citet{mor11}, from {\em Spitzer}/IRAC (3.6/4.5$\mu$m)
lightcurves  of ONC sources, with a much lower incidence,
$\sim$5\%, than found by \citet{ale10}, however.

In the X-ray band all PMS stars, both CTTS and WTTS, are intense sources
\citep{fei81,fla03b,pre05a}. Thermal emission from magnetically confined
plasma (i.e. coronae) at $\sim$10-30\,MK accounts for most of the
observed emission. Accreting stars (CTTS) also show a softer plasma
component at $\sim$2-4\,MK that, given its temperature and plasma
density, can be interpreted as the direct product of the accretion
shock. This component has only been unambiguously detected in the few
CTTS for which high-resolution X-ray spectra were obtained
\citep[e.g.][]{kas02,gue07b,arg11}, as demonstrated by emission-lines
from cool plasma and by density-sensitive He-like triplets. It is
natural to think that such a component is a common feature of CTTS.
However, for most stars, with the possible exception of TW~Hya
\citep{kas02,dup12}, this cool plasma does not contribute significantly
to the X-ray flux detected in the 0.3-10\,keV energy band covered by
present-day imaging detectors, such as those onboard of {\em Chandra}
and {\em XMM-Newton}. In fact, in spite of this additional soft
component, CTTS are, on average, two to three times fainter in X-rays
compared to WTTSs of the same mass or $L_{\rm bol}$
\citep{neu95a,dam95,fla03a,pre05a}. Moreover, their activity levels at
any given mass or $L_{\rm bol}$ show a larger scatter \citep{pre05a}.
These differences cannot be ascribed to differences in stellar rotation,
the main factor that determines activity levels on MS stars, and remain
largely unexplained. A higher average absorption of CTTS with respect to
WTTS, which would lower the observed X-ray fluxes of CTTS (but not the
intrinsic ones), was excluded by \citet{pre05a} on the basis of the {\em
Chandra} Orion Ultra-deep Project (COUP) data \citep{get05}, which
allowed determining individual absorption corrections for all stars.
These authors, as well as others, instead proposed that accretion might
modify the magnetic field structure \citep[and thus coronae, see
also][]{fla03b, tel07} or the stellar structure in such a way as to
influence the stellar dynamo. \citet{tel07} in particular, based on data
from the {\em XMM-Newton Extended Survey of the Taurus molecular cloud}
(XEST),  proposed that the accreting material from the disk {\em cools}
down part of the coronal plasma to temperatures outside the X-ray
regime. \citet{gre07} modeled coronal and accretion structures by
extrapolating magnetic field surface maps (from Zeeman-Doppler imaging)
and concluded that the lower observed X-ray luminosity of CTTS might
indeed be due to the obscuration of {\em large parts} of the corona by
dense gasseous accretion streams. This scenario is closely related to
that invoked to explain the AA Tau-like optical variability, the main
difference being that for X-rays the absorbing material would be the
gas in the accretion streams rather than the dust in the inner disk
warps at the bases of the same accretion steams.  The accretion streams
would only intercept a fraction of the emitting coronal plasma and the
emerging X-ray flux would thus be composed of two components, only one
of which is heavily absorbed. As also discussed in this paper, correctly
correcting the observed X-ray flux for extinction would be almost
impossible, at least with the spectral resolution and sensitivity of
present-day instruments.

In addition to the differences in observed X-ray flux, other differences
in the X-ray emission of CTTSs and WTTSs have been reported.
\citet{fla06}, for example, reported tentative evidence that CTTSs in the
NGC\,2264 star-forming regions are more X-ray variable than WTTS. The
same data, as well as the XEST data \citep{tel07}, also indicate that
the plasma on CTTSs is, on average, hotter than that on WTTS. At the
same time, some CTTSs show evidence for a plasma component cooler than
seen on WTTSs. While this latter observation may be attributed to plasma
heated in accretion shocks, the harder and more time-variable emission
could be consistent with time-variable absorption from circumstellar
material (accretion stream and/or disk warps).

From this discussion it is clear that studies of X-ray variability
may provide useful information on the mechanism that regulates either
the observed or the intrinsic X-ray activity level of PMS stars. All T
Tauri stars are strongly variable in the X-ray band on timescales
of hours and days: the most prominent form of variability is undoubtedly
represented by flares, with short rise phases during which the flux
reaches up to $\sim$100 times the pre-flare level, followed by slower
decay phases lasting several hours (see e.g. \citealt{fav05} for a
detailed study of bright flares in the COUP observation; \citealt{wol05}
and \citealt{car07} for statistical studies). Modulation due to the
inhomogeneity of coronal structures and stellar rotation has also been
detected by \citet{fla10} in the $\sim$13-days long COUP observation
(including gaps; the actual exposure time was $\sim$850ks). This work
also highlighted the difficulty in separating flaring from other
forms of variability. However, variability on timescales longer than one
day has generally not received much attention, partly because of the
scarcity of appropriate datasets. Moreover, the highly stochastic nature
of the X-ray emission has hindered statistical studies trying to
correlate the "average" properties of variability with stellar and
circumstellar characteristics.

An interesting result has recently been obtained by us from the
correlation of X-ray and optical emission. In a previous study,
\citet{sta07} had already searched in vain for a correlation between the
COUP X-ray lightcurves and simultaneous ground-based optical multi-band
lightcurves. \citet{fla10} have instead employed simultaneous
space-based observations of NGC\,2264 members by {\em Chandra} (X-rays)
and CoRoT (optical, broad-band). By comparing X-ray and optical fluxes
obtained in two 30\,ks time windows separated by 16 days, we have found
that, for CTTSs only, the X-ray variability in the soft 0.5-1.5\,keV
band is significantly correlated with the optical variability. This
correlation is seen neither for WTTS nor for the harder X-ray band
(1.5-8.0\,keV). We interpret these findings as caused by the effect of
time-variable absorption from circumstellar material. From the relative
magnitude of the X-ray and optical variability we conclude, moreover,
that the absorbing material must be dust-depleted, thus pointing toward
accretion streams within the dust-sublimation radius, as proposed by
\citet{gre07}.

If the X-ray emission of CTTSs is indeed absorbed by circumstellar
structures in co-rotation with the star, their X-ray variability should
$i)$ be more extreme than on WTTS, $ii)$ reach maximum amplitudes on the
timescale of the stellar rotation, and $iii)$ show clearer signs of
rotational modulation. Most X-ray observation of nearby young stars
performed with present-day instruments are shorter than $\sim$100ksec
($\sim$1d) and therefore unsuitable for studying variability on the
timescale of typical stellar rotations \citep[$\sim$1-15\,d for stars in
the ONC;][]{her02}. The most notable exception is the almost
uninterrupted 850\,ks long COUP observation of the ONC. We have thus
decided to exploit this dataset once more to investigate X-ray
variability as a function of timescale from hours to $\sim$12 days, thus
covering the stellar rotation of most ONC stars. In addition to its
length, the COUP observation with its $\sim$1600 X-ray sources also
provides the advantage of a rich and optically well-characterized sample
of PMS stars. Given the highly stochastic nature of variability, large
stellar samples are crucial to reliably determine "average" variability
properties for different stellar subsamples. We can then try to
correlate these average properties with stellar and circumstellar
characteristics to investigate the physical origin of variability.

This paper is organized as follows: \S\,\ref{sect:data} describes the
COUP dataset; \S\,\ref{sect:analysis} illustrates the two methods we
used to quantify variability, one non-parametric
(\S\,\ref{sect:np_analysis}) and one tailored to impulsive flares
(\S\,\ref{sect:flare_analysis});\S\,\ref{sect:samples} describes the
selection of the stellar samples used for the analysis.
Section\,\ref{sect:results} presents the result of the variability
analyses, which are then more fully discussed in
\S\,\ref{sect:discussion}, along with their implications.  Finally, we
summarize our main conclusions in \S\,\ref{sect:summconc}.

\section{The COUP data}
\label{sect:data}

The COUP observation is described in detail by \citet{get05} along with
the procedures followed for data preparation, source detection,
identifications with optical/NIR catalogs, and photon extraction. We
used the products of this analysis and, in particular, the original
lists of source and background photons extracted for each source. We
also adopted from \citet{get05} the stellar and circumstellar parameters
for the COUP sources, both collected from the literature or derived by
these authors (e.g. magnitudes, masses, etc.). We updated this database
with respect to stellar rotational periods and classifications in terms
of near-IR excesses indicative of disks and envelopes. We used the same
collection of rotational periods as \citet{fla05}, with the addition of
48 new periods derived from an $\sim$250\,h {\em Spitzer} observation of
the ONC performed as part of the "Young Stellar Object VARiability"
(YSOVAR) program and reported by \cite{mor11}. From this work we
also adopted the classification in terms of spectral energy distribution
(i.e. Class\,I, Class\,II, and Class\,III).

\section{Analysis methods}
\label{sect:analysis}

We first discuss the non-parametric variability analysis that is at the
basis of our following discussion, and then the method we used to
asses the contribution of X-ray flares to the total source variability.

\subsection{Variability amplitudes}
\label{sect:np_analysis}

\begin{figure}[!t!]
\centering
\includegraphics[width=9.0cm]{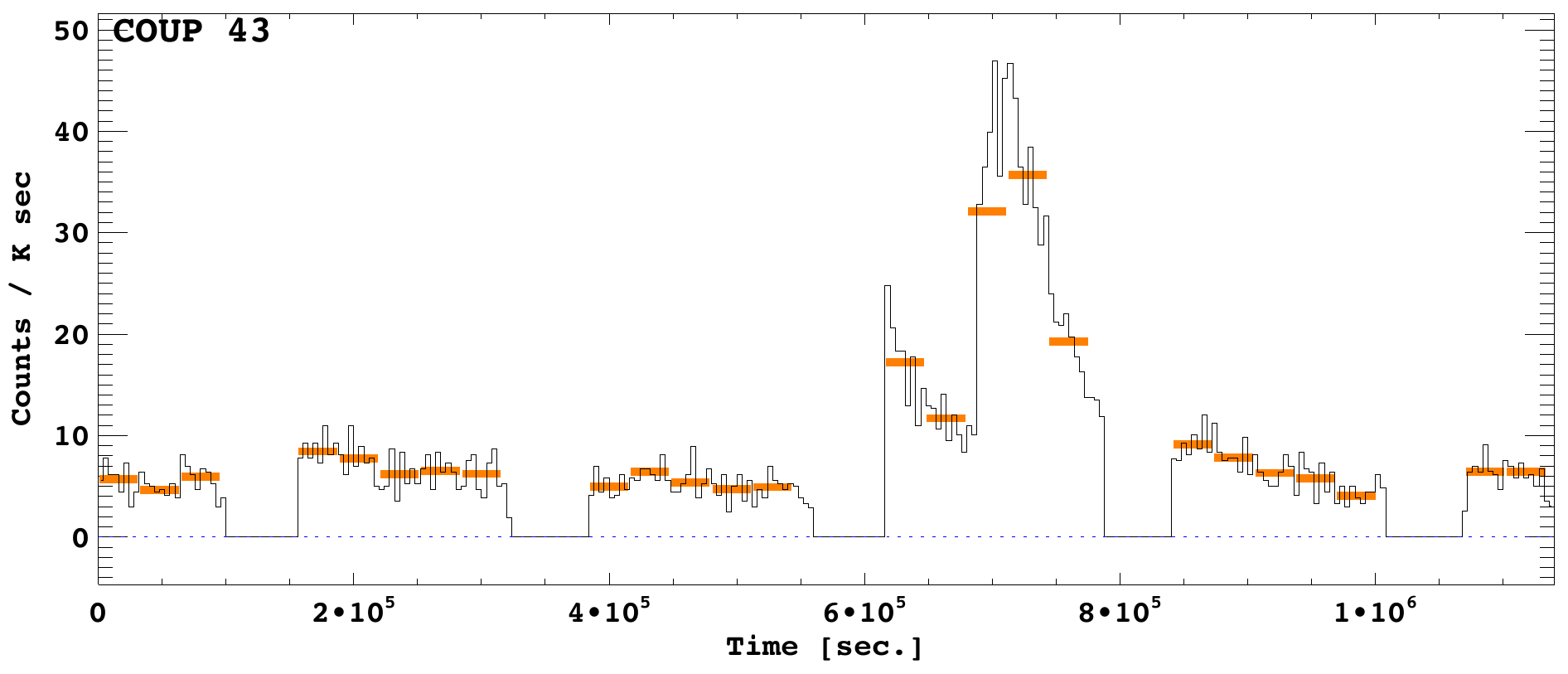}
\caption{Example X-ray light-curve for a bright COUP source. The
horizontal segments, superimposed on the lightcurve (black line),
indicate time intervals and relative photon fluxes for the 25 30\,ksec segments on which the variability analysis is based.}
\label{fig:exLC}
\end{figure}

To study how the X-ray fluxes vary as a function of timescale, we
subdivided the 850\,ks COUP observation into shorter segments: we chose
to adopt segment lengths of 30\,ks, as a compromise between being able
to probe short timescales, specifically with respect to the typical
stellar rotation periods,  and the need to have a sufficient number of
detected photons per segment and thus limit Poisson uncertainties. With
this choice we selected 25 time segments within the COUP observation,
for each of which we measured the background-subtracted photon flux.
Figure\,\ref{fig:exLC} illustrates, superimposed on a binned light-curve
(E=0.5-8.0\,keV) of one of the brighter COUP sources, the time segments
we adopted for the following analysis.

In general terms, our analysis consists of quantifying the variability
between segments separated by a given time interval, $\Delta t$, by
taking the ratio between the fluxes in the first and second segment,
$F_1/F_2\equiv F(t)/F(t+\Delta t)$\footnote{We experimented with
different ways to measure variability between segments, all yielding
consistent results.}. We then built distributions of these variability
measures (i.e. flux ratios) for all segment pairs within the same
lightcurve separated by $\Delta t$ and for all X-ray sources in given
subsamples. We then study how the {\em width} of the distributions of
variability measures obtained for the same subsample varies as a
function of timescale, $\Delta t$, and compare the trends for different
subsamples\footnote{The number of datapoints (flux ratios) that define
the distributions is obviously dependent on timescale, being generally
lower at longer $\Delta t$.}. For our purposes the width of the
distribution is defined as {\em the difference between the 75\% and the
25\% quantiles}. We refer to this width as the {\em variability
amplitude} ($VA$) of the sample at the given $\Delta t$. For stars for
which we also know the rotation period, we also explore the dependence
of VAs on rotational phase differences. To this purpose we construct and
compare the widths of distributions of flux ratios taken between
segments separated by rotational phase differences: $\Delta \phi= \Delta
t/P_{rot}$.

\begin{figure}[!t!]
\centering
\includegraphics[width=8.5cm]{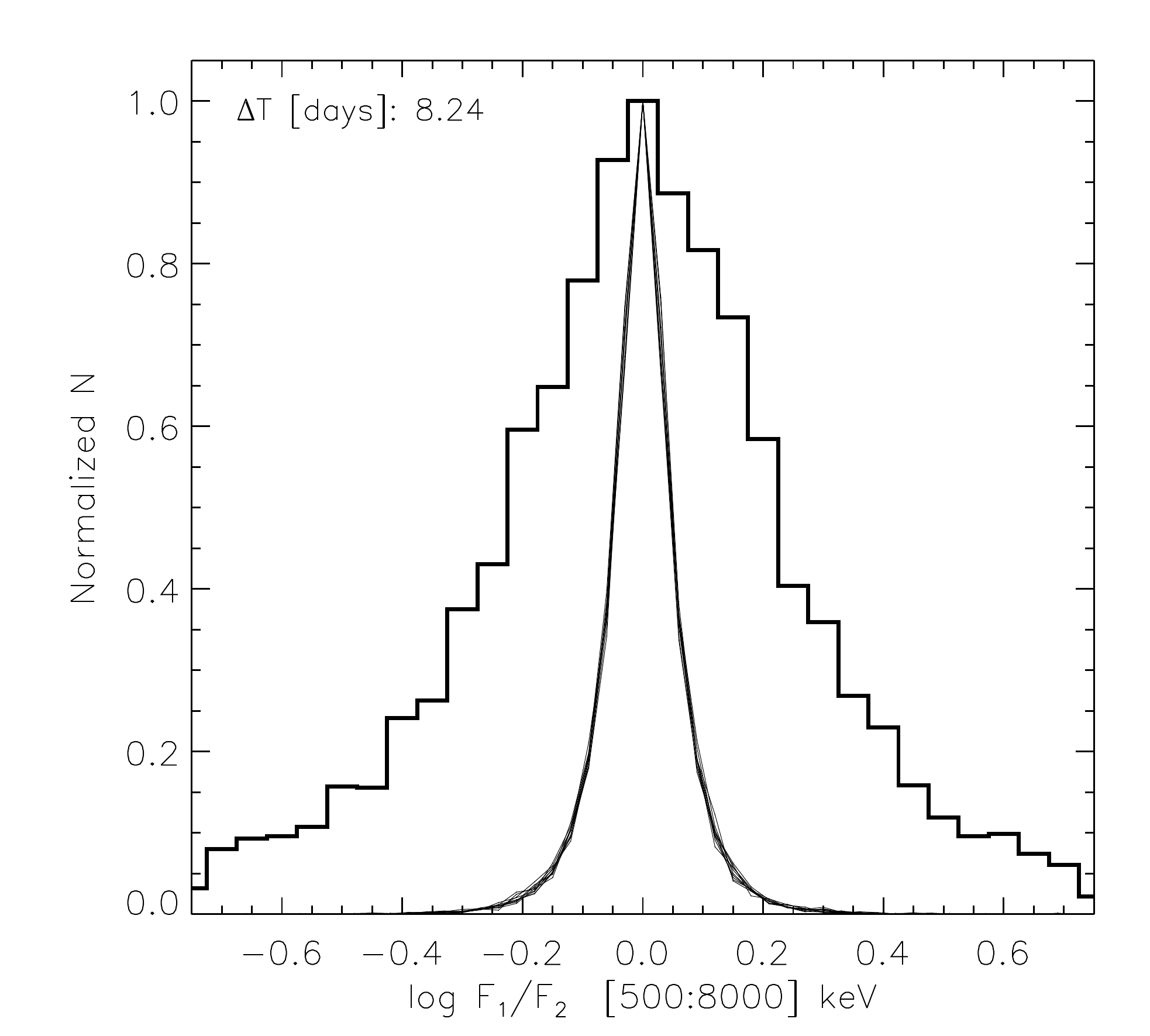}
\caption{Normalized histogram of all flux ratios, $\log(F_1/F_2)$, 
taken between 30\,ksec time intervals separated by  $\Delta t =$7.1-9.6
days (logarithmic mean: 8.24 days), for the 167 Class\,III stars in our
flux-limited sample (broken line). The narrower distribution, actually 
a bundle of ten almost indistinguishable ones, represents the contribution 
of Poisson statistics to the
observed distribution as determined from Monte Carlo simulations (see
text).}
\label{fig:FxRatioDist}
\end{figure}

The whole analysis was carried out considering three energy bands: the
full {\em Chandra} band (0.5-8.0\,keV), a soft band (0.5-1.5\,keV), and
a hard band (1.5-8.0\,keV). As an example, we plot with a thick broken
line in Fig.\,\ref{fig:FxRatioDist} the differential distributions of
$\log(F_1/F_2)$ obtained for the {\em full} X-ray spectral range and for
$\Delta t \sim$8 days (effectively, 7.1$<\Delta t<$9.6 days). In this
example the stellar sample comprised 167 Class III stars (see below for
sample definitions). Given the adopted $\Delta t$ range, each of these
stars contributes 47 flux ratios and the distribution is defined by 7849
datapoints.

The finite width of the $\log(F_1/F_2)$ distributions, i.e. their
departure from $\log(F_1/F_2)=0.0$, can be caused by two factors: real
flux variability, or Poisson uncertainties in the number of detected
photons. To isolate this last unwanted contribution, we have taken two
approaches: first, we limited our search sample to X-ray sources with
``good'' statistics; second, we quantified the effect of Poisson noise
on the variability distributions through Monte Carlo simulations. More
specifically, we imposed a limit of 750 counts within the 25 chosen time
segments, i.e. we required $>$30 counts per segment (or 1 count/ksec)
{\em on average}. This condition reduces the number of available X-ray
sources from the initial 1616 to 496, 319, and 361 for full, soft, and
hard band, respectively. As discussed in more detail in
\S\,\ref{sect:discPhyImp}, this selection introduces a bias on the
resulting samples, quite directly in terms of $L_X$, and indirectly,
through the correlation of $L_X$ with fundamental stellar properties
\citep[e.g.][]{pre05a}, in terms of stellar mass and bolometric
luminosity.

To evaluate the effect of  Poisson uncertainties on the flux-ratio
distributions, and ultimately on the $VA$s, we generated for each of the
considered distributions (i.e. for each sample and timescale) 1000
simulated flux-ratio distributions assuming that sources are
intrinsically constant and that the measured fluxes only vary because of
Poisson fluctuations.\footnote{More specifically, for each pair of time
segments involved in the flux-ratio distribution we generated simulated
source (and background) fluxes, from Poisson distribution appropriate
for a flux equal to the mean of the two measured fluxes.} Figure
\ref{fig:FxRatioDist} shows ten of the simulated constant-source
distributions (the bundle of narrow distributions centered at
$\log(F_1/F_2)=0.0$). In this case it is quite clear that Poisson
uncertainties cannot account for the widths of the actual flux-ratio
distributions. In general, we quantified the contribution of Poisson
statistics on the $VA$ of each of the considered samples and timescales
by recording the medians and $1\sigma$ dispersions of the $VA$s measured
from the 1000 simulated constant-source fluxes.

\subsection{Flare analysis}
\label{sect:flare_analysis}

Coronal flares, characterized by short rise-phases and slower decays,
are an obvious source of variability for our sources. To discuss the
contribution of flares to the overall variability amplitudes
(\S,\ref{sect:disc_flares})  we  seek to quantify flaring activity
statistically for a given stellar sample.  X-ray flares in the COUP
datasets have already been investigated by at least four studies
\citep{fav05,wol05,car07,joh12}. We adopted the technique used by
\citet{wol05} and \citet{car07}, that is, we detected individual flares
using an automated procedure and then derived their rate of occurrence
as a function of their intensities as well as the distribution of
``flare durations''. A full description of the method can be found in
\citet{car07}.

\section{Sample definitions and selection criteria}
\label{sect:samples}

As described in the previous section, our parent samples for the three
X-ray spectral bands comprise all COUP sources with an average
count-rate $>$1 count/ksec in the 25$\times$30\,ks selected time
intervals.  We then selected physically relevant and homogeneous
stellar samples for our study of variability as a function of timescale
and stellar characteristics.

First, we excluded high-mass stars, $\rm M>3M_\odot$, the X-ray emission
of which is of non-coronal origin, or comes from a different kind of
corona, or originates in unresolved lower mass companions
\citep[e.g.][]{dam94,zin94,ham05,ste09}. Then, to further restrict the
range of masses, and that of absorptions, we excluded stars with $\rm
I_c>$16,  corresponding to $\rm M > 0.1-0.2 M_\odot$ for COUP sources
with low optical absorption\footnote{For the low-mass limit we chose to
rely on the $I_c$ magnitude instead of the masses estimated from the HR
digram and evolutionary models. This is because this latter
determination of masses, ultimately based on spectral types, is less
complete at the low-mass end. We would thus have lost a number of
otherwise useful stars for the following analysis.} ($\rm A_V < 2.0$).
Finally we excluded proper-motion non-members according to the
collection of \citet{hil97}, unless they had indications of
circumstellar accretion or circumstellar material (see
below). With these selection our search samples are reduced to 369, 289,
and 261 stars for the full, soft, and hard bands, respectively.

Next, we defined the following subsamples of low-mass stars based on
their accretion properties, presence of circumstellar disks, and mass:

\begin{itemize}
\item CTTS and WTTS, based on the equivalent width (EW) of the
8542\,$\AA$ Ca\,II line, following the same approach as
\citet{fla03b} and \cite{pre05a}. Namely, we selected as accreting stars
(CTTS) those with the Ca II line in strong emission, EW$<$-1, and as non-accreting stars (WTTS) those whose Ca II line clearly is in absoption: EW$>$1. By excluding intermediate EW values, this approach results in better
defined and {\em extreme}  samples: the selected CTTS, in particular,
will be biased toward high accretion rates.

\item Class II and Class III, based on the observation of  mIR excesses
due to warm circumstellar material, as reported by \citet{mor11}.

\item Low- and intermediate-mass Class II and Class III stars, defined as
complementing the above selections with the following constraints on
mass:  $\rm M < 0.5 M_\odot$ and  $\rm M>0.8 M_\odot$. 
\end{itemize}

We considered other possible subsamples, but, since their sizes are too
small to derive meaningful results, we will not discuss them. For the
samples and X-ray energy bands discussed in the following,
Table\,\ref{tab:samples} summarizes the size of the sample and, if
relevant, the size of the subsample of stars for which we know the
rotational period. Most of our analysis will be based on the  samples of
CTTSs, WTTSs, Class\,II, and Class\,III stars, which we therefore refer
to as our main samples.\footnote{There is significant but not complete
overlap between CTTS and Class\,II stars and between WTTS and
Class\,IIIs. In particular, among the 57 CTTSs, 2 are classified as a
Class\,I, 31 as Class\,II, 14 as Class\,III and 10 have no nIR
classification; among the 110 WTTSs, one is a Class\,I, 29 are
Class\,II, 70 are Class\,III, and 10 are have no nIR classification.}

\begin{figure}[!t!]
\centering
\includegraphics[width=9.0cm]{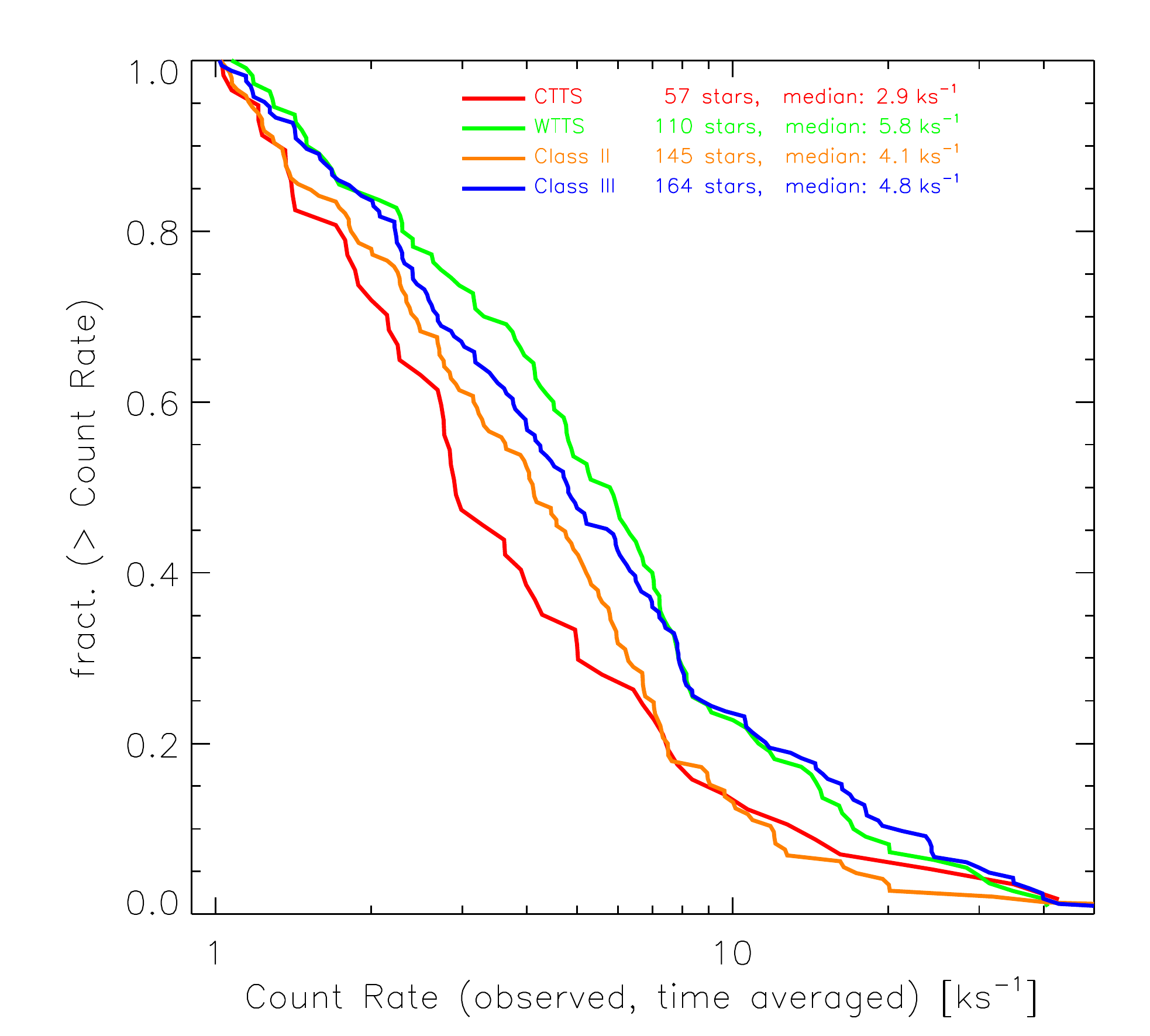}
\caption{Distributions of count rates for our flux-limited samples of CTTS (red), WTTS (green), Class\,II (orange), and Class\,III (blue) stars. Sample sizes and median count rates are reported in the legend. Count rates are  defined here as the total number of counts detected in the 25$\times30$\,ksec time intervals under consideration, divided by the exposure time, 750\,ksec.}
\label{fig:crate_dist}
\end{figure}

In the following we used of the statistical properties of the X-ray
sources in our samples to constrain the origin of their X-ray
variability. Figure\,\ref{fig:crate_dist} shows the distribution of
(full-band) count-rates for our main samples. In spite of our choice of
count-rate limited samples, and consistent with the results of
\citet{fla03} and \citet{pre05a}, CTTS have statistically lower
count-rates with respect to WTTS (a factor of $\sim$2 in the median). A
difference is also seen between Class\,II and Class\,III, but it is
smaller, possibly owing to the less extreme nature of the selection
criteria. Significant differences in median count-rates are also
observed between  low- and intermediate-mass Class\,II and Class\,III
stars, with the higher mass samples being systematically brighter than
the low-mass ones by a factor of $\sim$2 and $\sim$3 for Class\,II and
Class\,III, respectively.

\begin{table}[] 
\caption{Sizes of the analyzed samples. Figures in parentheses refer to subsamples with known rotational periods} 
\label{tab:samples}
\centering 
\begin{tabular}{l |r r r | r | r } 
\hline\hline 
\multirow{2}{*}{Sample} & \multicolumn{3}{|c|}{All masses}   & $<$0.5\,$\rm M_\odot$ & $>$0.8\,$\rm M_\odot$\\  
			& 					{\scriptsize 0.5-8\,kev}	 & 	{\scriptsize	0.5-1.5\,kev} 		&     		{\scriptsize	1.5-8\,kev} & 	{\scriptsize	0.5-8\,kev}	  		& {\scriptsize 0.5-8\,kev}  \\
\hline
CTTS		&             57	 (19)  &           33 (11)	& 		41 (13) 	&        - 	   		&    -     	   	\\
WTTS	&           110	 (72)  &           93 (62)	&	 	82 (53)		&         -	   		&     -    	   	\\
Class II	&           145	 (70)  &         109 (54) 	& 	  109 (52)	&    	 55     	   	&   27      	   	\\
Class III	&           164 (105) &         138 (94)	&    113 (73) 	&    	 73    	   	&   43           \\
\hline   	
\end{tabular} 
\end{table}

\section{Results}
\label{sect:results}

For the stellar samples introduced in \S\,\ref{sect:samples} we first
present the results of our non-parametric analysis of variability
amplitudes as a function of timescale and rotational phase. Next, as a
tool to constrain the origin of variability, we investigate the
incidence of flaring activity on the same samples.

\subsection{Variability amplitudes}
\label{sect:results_VA}

As described in \S\,\ref{sect:analysis}, we defined the representative
variability amplitude, $VA(\Delta t)$, of a  given source sample and for
a given timescale $\Delta t$, as  the 50\% width (difference between
75\% and 25\% quantiles) of the distribution of all the $F(t)/F(t+\Delta
t)$ that can be computed from the X-ray light curves of stars in that
sample.  We now discuss how this quantity depends on $\Delta t$ and
on stellar characteristics by comparing different source samples. Next,
we repeat the same comparative analysis as a function of stellar
rotational phase (instead of timescale) by normalizing the time-lag
$\Delta t$ by the  stellar rotation period, $P_{rot}$, i.e. by
considering $VA(\Delta \phi)\equiv VA(\Delta t/P_{rot})$.

\begin{figure*}[]
\centering
\includegraphics[width=8.4cm]{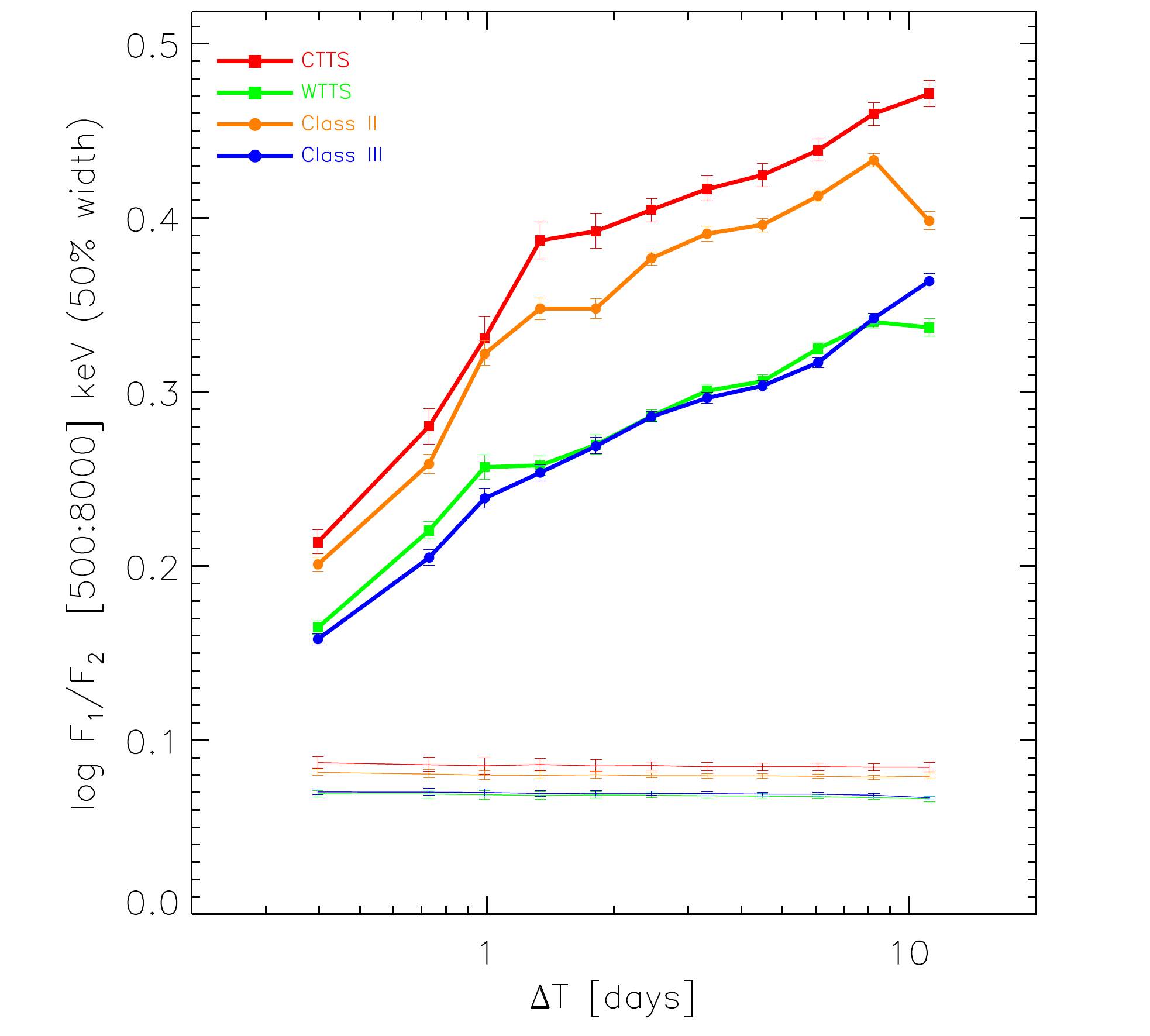}
\includegraphics[width=8.4cm]{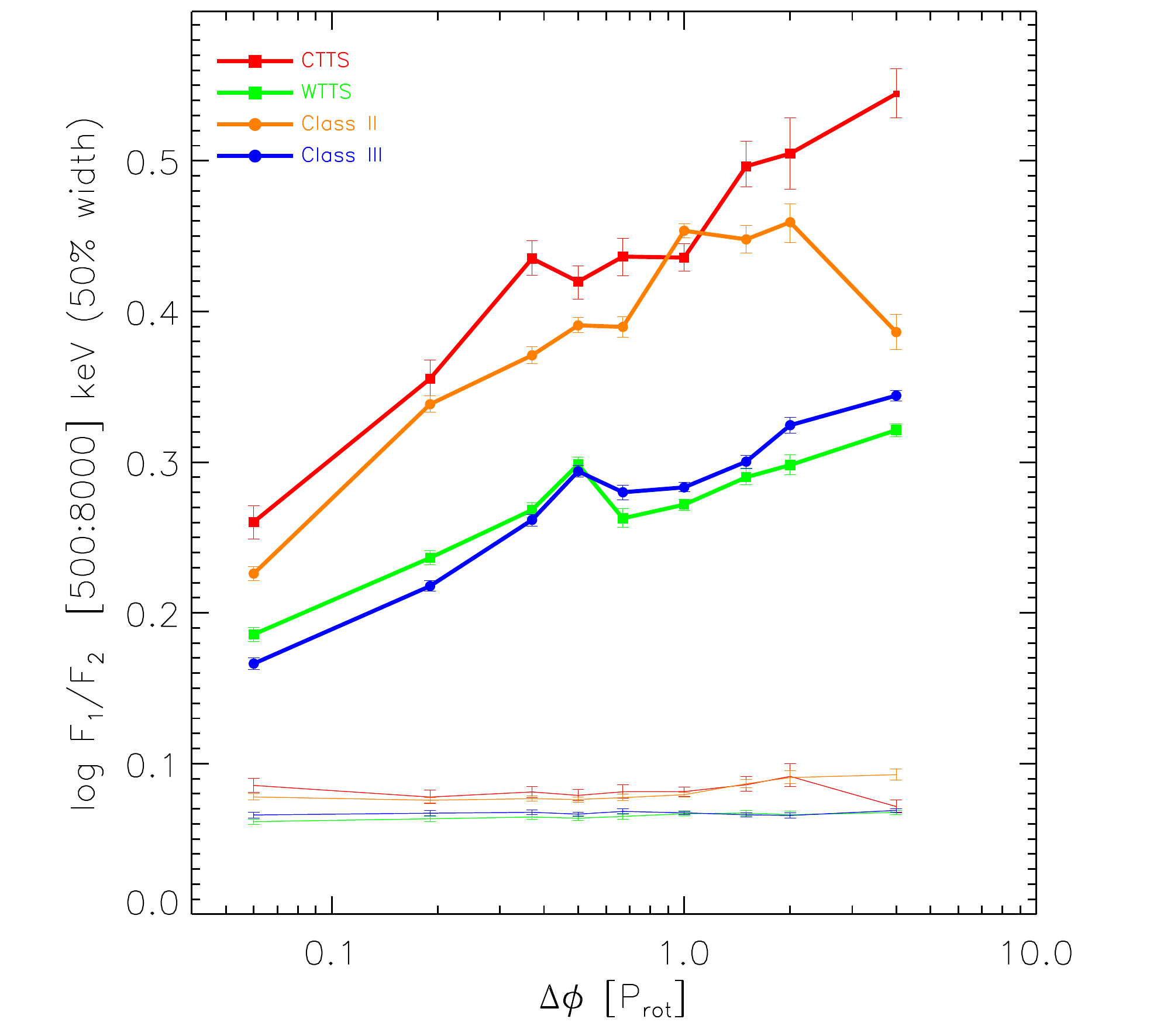} \\
\includegraphics[width=8.4cm]{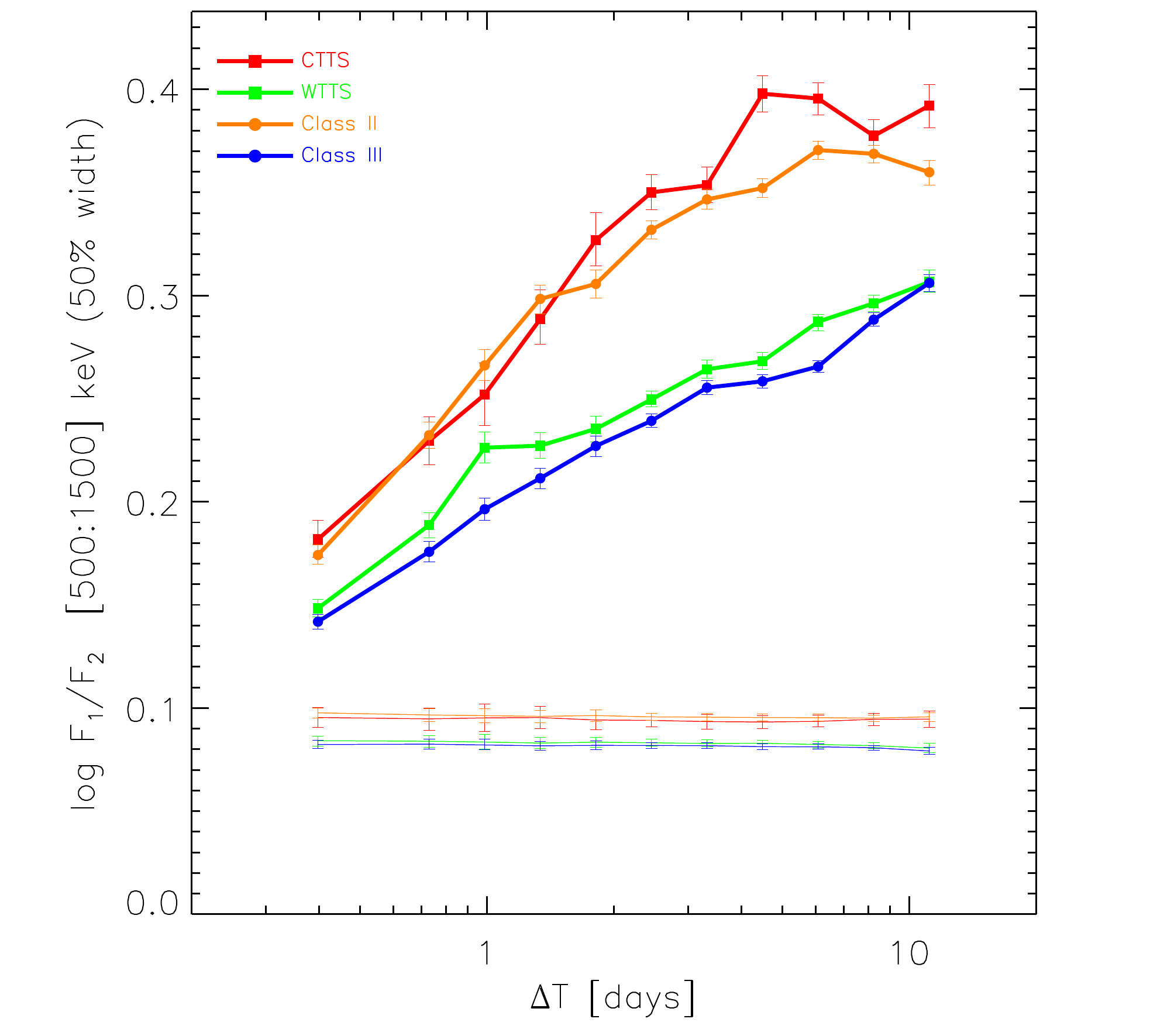} 
\includegraphics[width=8.4cm]{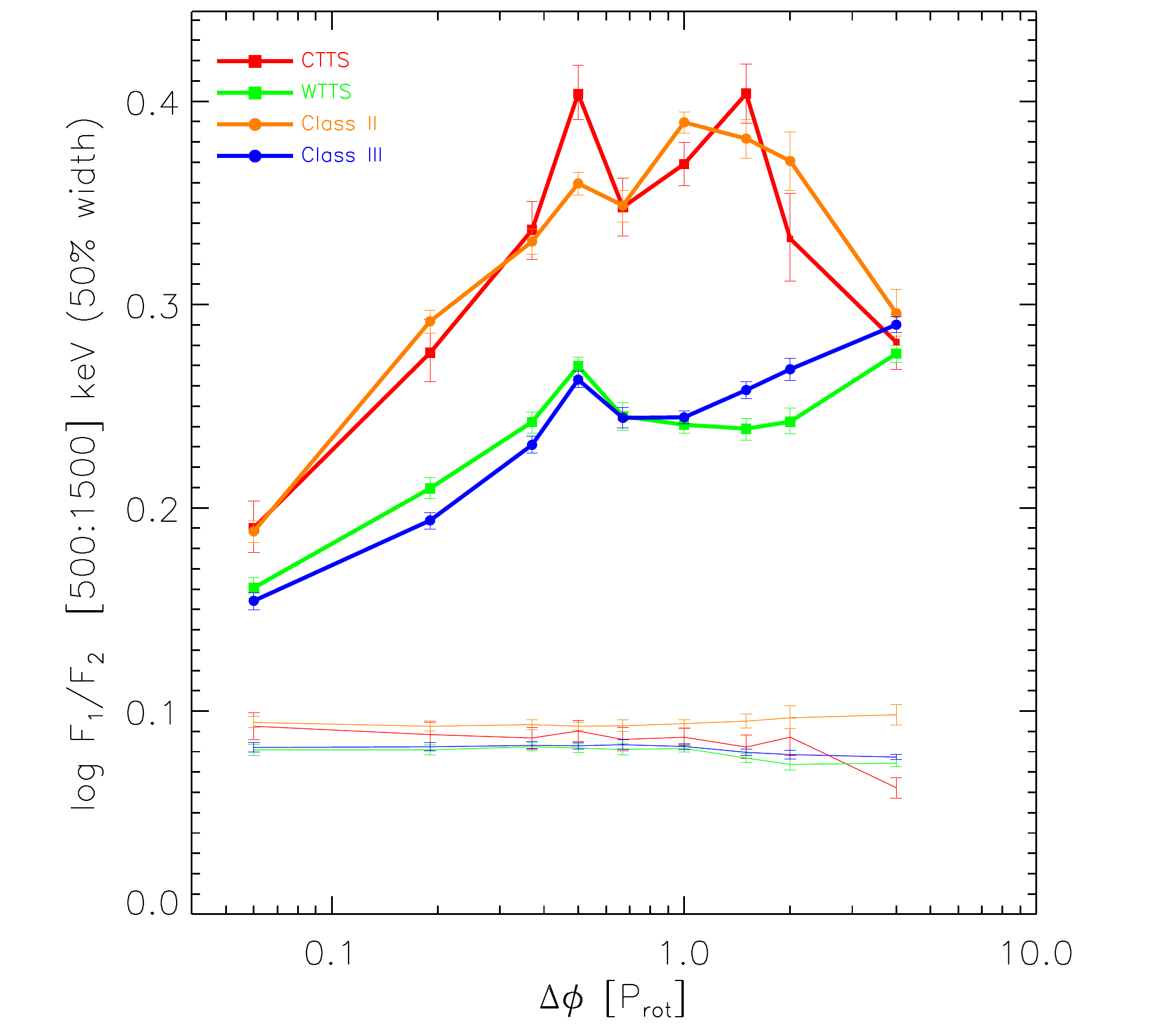} \\
\includegraphics[width=8.4cm]{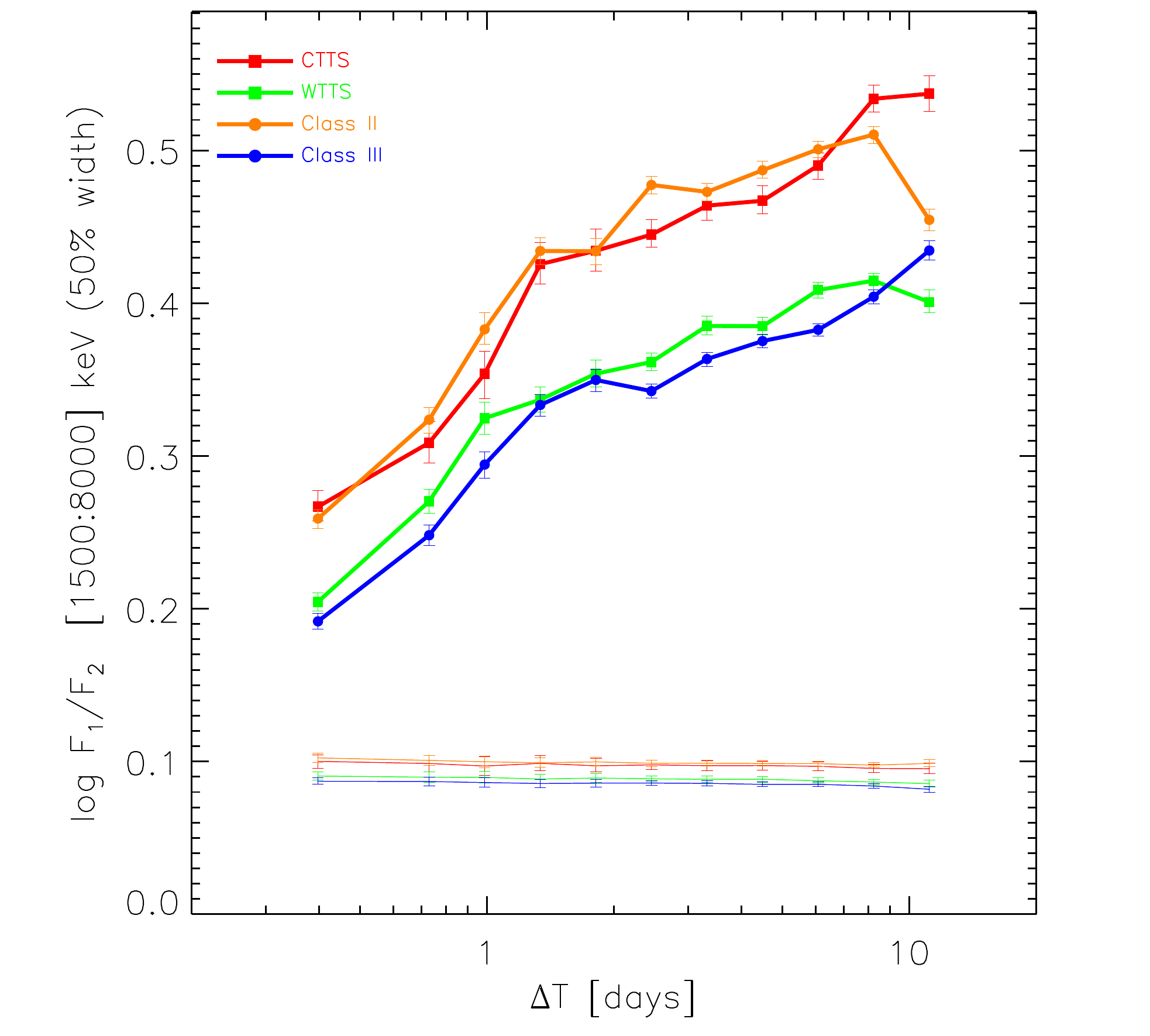}
\includegraphics[width=8.4cm]{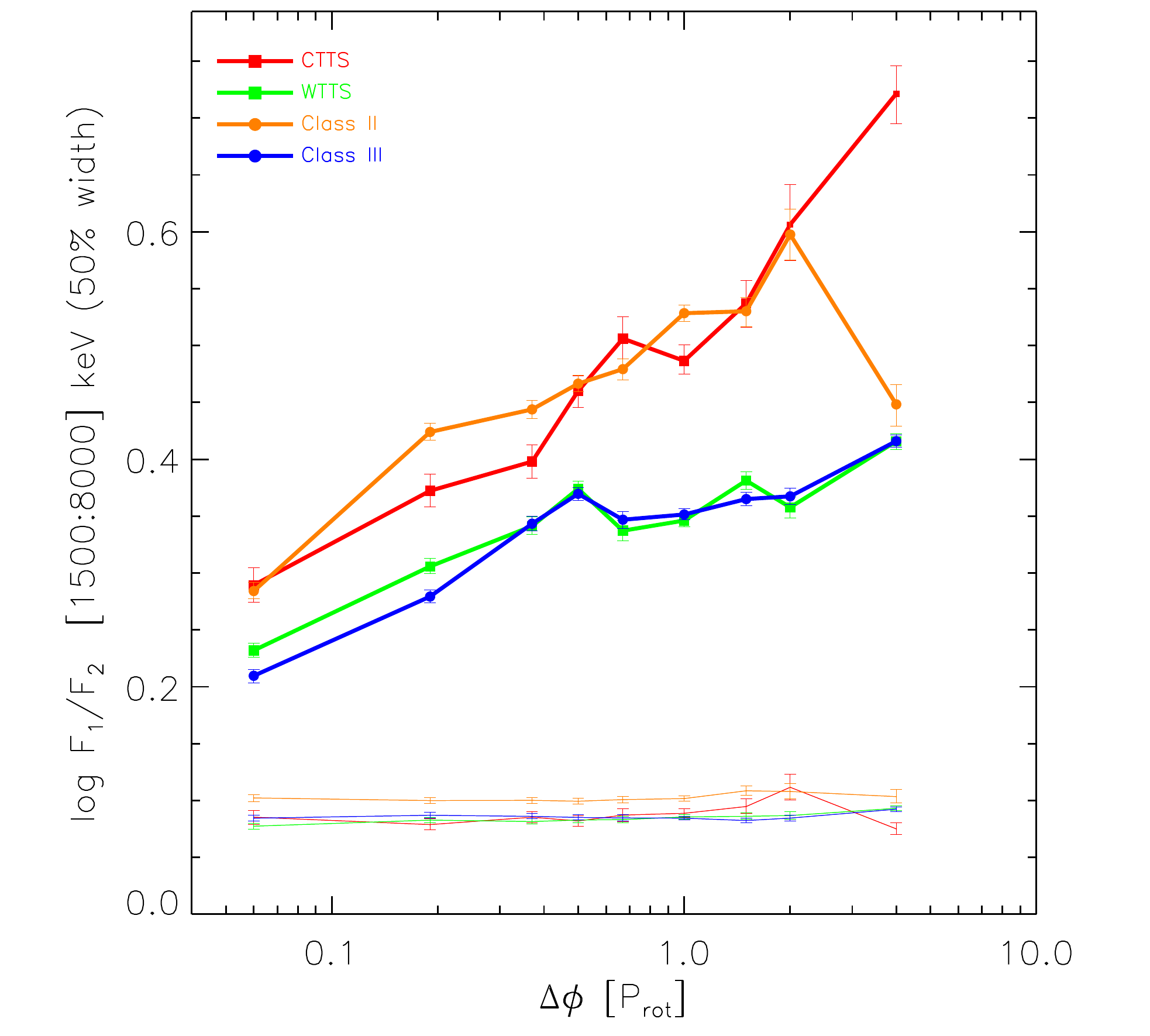} \\
\caption{Variability amplitude, $VA$, as a function of timescale (left
column) and rotational phase difference (right column), for CTTS, WTTS,
Class\,II, and Class\,III stars (see legends within panels), and for
three different X-ray energy bands: full (0.8-8.0\,keV, top row), soft
(0.5-1.5\,keV, middle), and hard (1.5-8.0\,keV, bottom). As described in
\S\,\ref{sect:analysis}, the $VA$ for each sample and x-axis value
($\Delta t$ or $\Delta \phi$)  is defined as the 50\% width of the
distribution of $\log F_1/F_2$ (cf. Fig.\,\ref{fig:FxRatioDist}), i.e.
of the difference between the 75\% and 25\% quantiles. Formal error bars reflect 
photon statistics and are estimated thorough Monte Carlo simulations.
Thick lines refer
to the VAs derived from the data, while thin lines with error bars and
matching color indicate the contribution of Poisson noise (also estimated 
with Monte Carlo simulations).}
\label{fig:VarAmp}
\end{figure*}

\subsubsection{Timescale}
\label{sect:results_VA_DT}

The left-hand column of Fig.\,\ref{fig:VarAmp} shows the run of $VA$ vs.
timescale in the interval  0.4-11\,days for CTTS, WTTSs, Class\,II,
and Class\,III stars.  The three rows refer to the three X-ray energy
bands: full, soft, and hard (from top to bottom). The contribution of
Poisson statistics to the $VA$s, as determined from Monte Carlo
simulations, is also shown (both median and $1\sigma$ dispersion).
Note that this latter contribution is not to be subtracted linearly from
the observed $VA$s. In the hypothesis the distributions of $\log
F_1/F_2$ are Gaussian (which is not far from being true, cf.
Fig.\,\ref{fig:FxRatioDist}), the Poisson contribution should be
subtracted in quadrature. They thus contribute little to most of the
measured $VA$s. The left-hand panels in Fig.\,\ref{fig:VarAmp} lead to
the following conclusions:

\begin{enumerate}

\item All samples are significantly variable on all timescales.

\item Variability ($VA$s) increases strongly with timescale, up to at
least 11\,days, i.e. the longest timescale probed by the COUP
observation. However, for some samples and energy bands (in particular
CTTSs in the soft band), $VA$s may saturate at long timescales.

\item CTTS and Class\,II stars appear to be systematically more
variable than WTTS and Class\,III stars.

\item The difference in $VA$ between CTTS/Class\,II and
WTTS/Class\,III stars seems to increase toward long timescales. This is
especially true for CTTSs and WTTSs in the soft band. 
\end{enumerate}

\begin{figure}[!t!]
\centering
\includegraphics[width=9.0cm,clip]{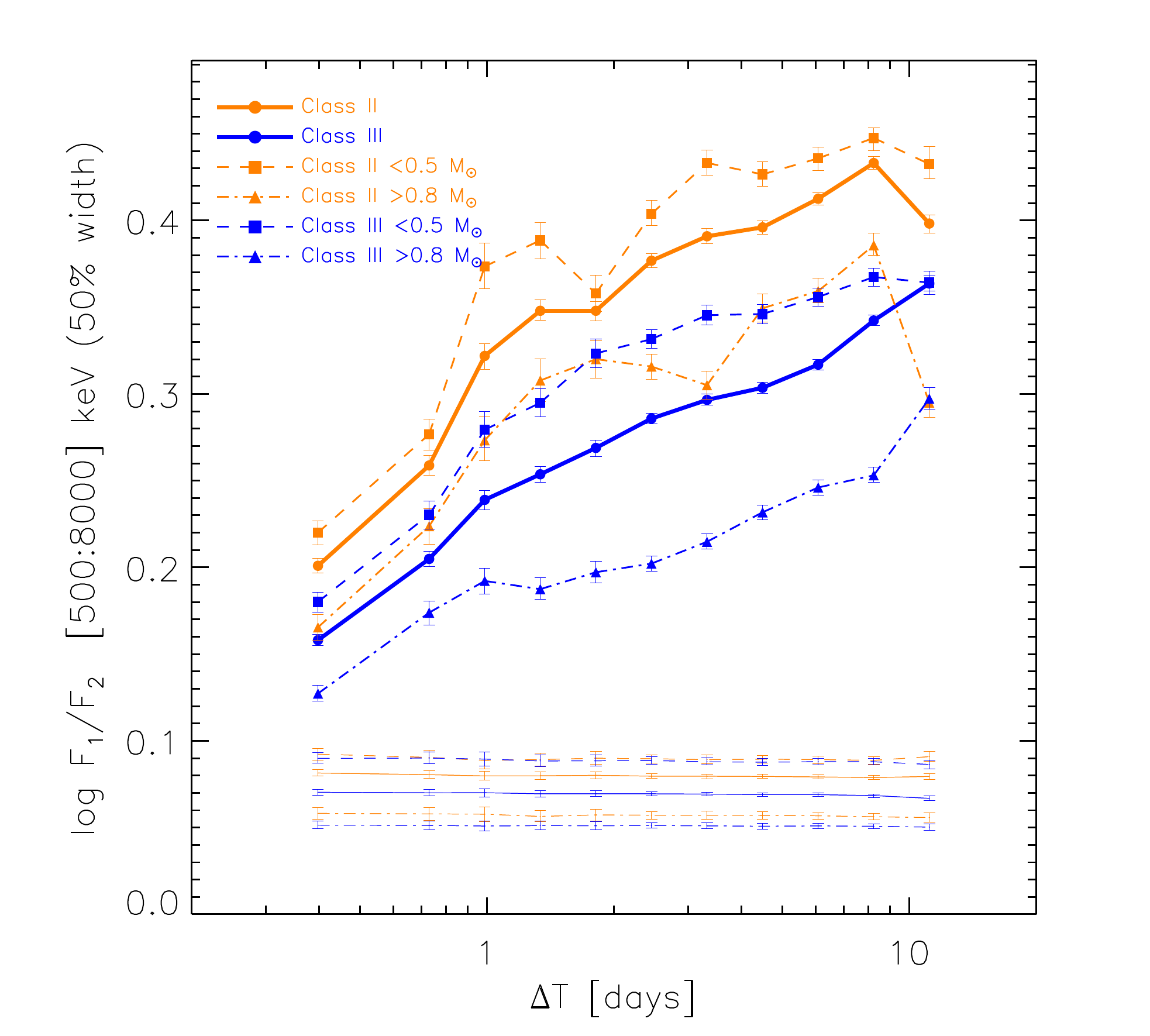}
\caption{Variability amplitude, $VA$, as a function of timescale for
Class\,II and Class\,III, for the full samples (same as in
Fig.\,\ref{fig:VarAmp}) and for two mass-segregated subsamples (see
caption).}
\label{fig:VarAmp_mass}
\end{figure}

We also studied the dependence of variability on stellar mass. Figure
\ref{fig:VarAmp_mass} shows again, repeated from Fig.\,\ref{fig:VarAmp},
the run of $VA$ vs. timescale for the two richest samples, Class\,II
and Class\,III stars, along with the trends obtained for the two
mass-segregated subsamples of stars in the same classes. We see that,
while the trends for the mass-segregated subsamples are qualitatively
the same as for the full samples, low-mass stars of both classes are
more variable than their higher mass counterparts.

\subsubsection{Rotational phase}
\label{sect:results_VA_RP}

The left-hand column of Figure\,\ref{fig:VarAmp} shows the run of $VA$
vs. the rotational-phase difference $\Delta\phi$, ranging from 0.05 to
4.0\,$P_{rot}$. The purpose of these plots is to investigate the
presence of rotational modulation, as evidenced by peaks at
$\Delta\phi=0.5$\,$\rm P_{rot}$ and $\Delta\phi=1.5$\,$\rm P_{rot}$.
Other possible peaks at higher $\Delta\phi$ values (e.g. 2.5) cannot be
investigated with our data since in order to have significant number of
stars in each $\Delta\phi$ bin we had to choose progressively wider bins
with increasing  $\Delta\phi$. We also note that, in contrast to the case
of the $VA(\Delta t)$ plotted on the left, the number of stars entering
in the definition of the $VA(\Delta\phi)$, i.e. those that define the
distributions of $\log F(\phi)/F(\phi+\Delta\phi)$, generally decreases
with $\Delta\phi$\footnote{For both $VA(\Delta t)$ and $VA(\Delta\phi)$
the number of usable flux-ratios {\em per star} varies along the
x-axis.}. For example, given the fixed length of our X-ray observation,
the variability at higher $\Delta\phi$ values can only be probed in
stars with shorter rotational periods. From these plots we draw the
following conclusions:

\begin{enumerate}
\item Several samples show, in one or more spectral bands, signs of
rotational modulation. In particular small but statistically significant
peaks at $\Delta\phi=1.5$ and/or 0.5\,$\rm P_{rot}$ are observed in all
three bands for WTTS and Class\,III stars.

\item Less clear signs of rotational modulations are generally seen for
CTTS and Class\,II stars, {\em except} in the soft band, and in
particular for CTTSs, for which a clear modulation signal is
revealed by pronounced peaks at both 0.5 and 1.5\,$\rm P_{rot}$.
The peaks are significant at the 3.8 and 2.3$\sigma$ level, when
compared to the mean of the two adjacent points\footnote{The
significance of the peak at 1.5\,$\rm P_{rot}$ would be higher if
computed with respect to the minimum at 0.75\,$\rm P_{rot}$.} and
suggest apparently larger amplitudes with respect to
WTTSs/Class\,III stars.
\end{enumerate}

\subsection{Flare frequencies and lengths}
\label{sect:flare_freq}

\begin{figure*}[!th!]
\centering
\includegraphics[width=9.0cm]{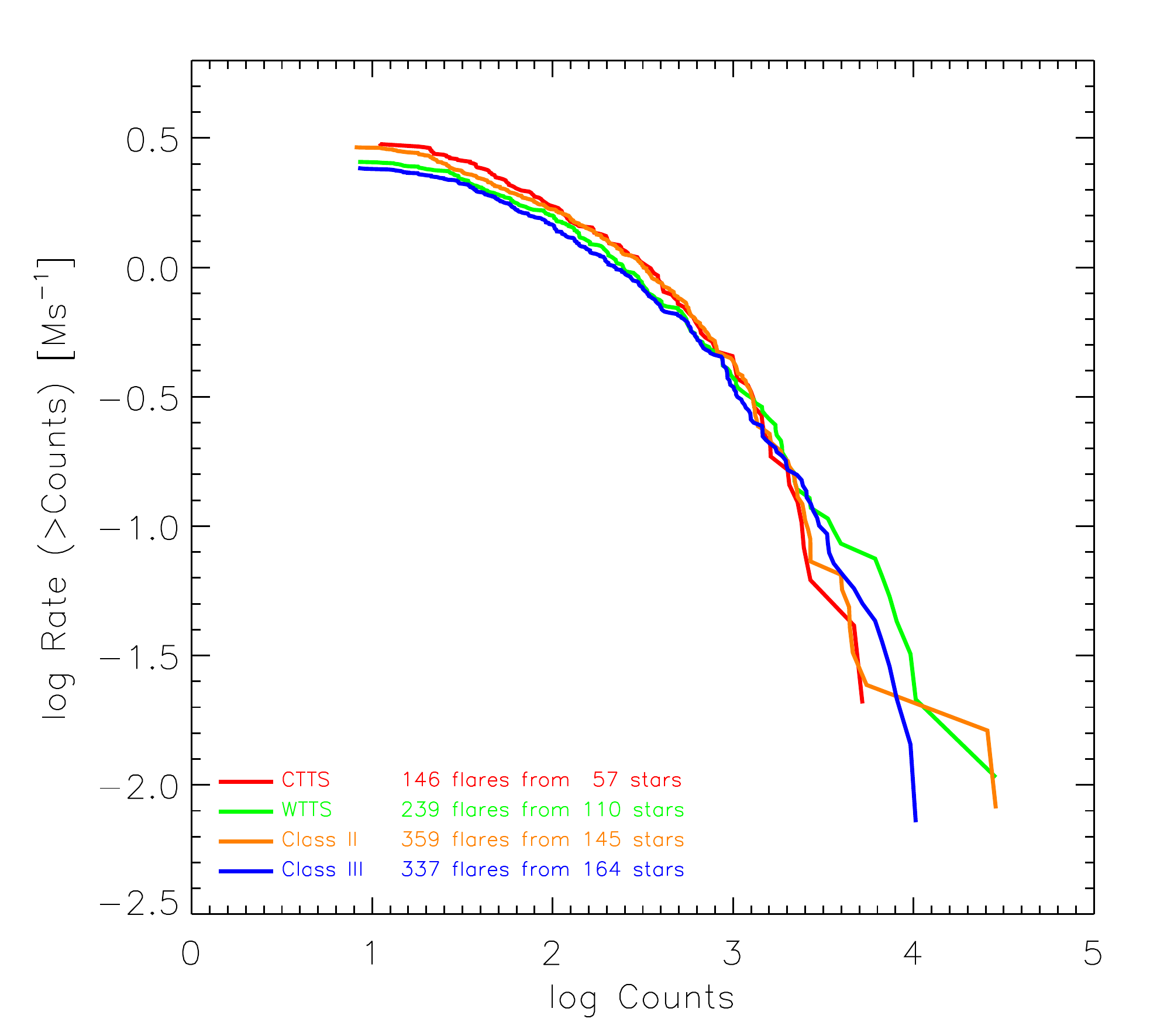}
\includegraphics[width=9.0cm]{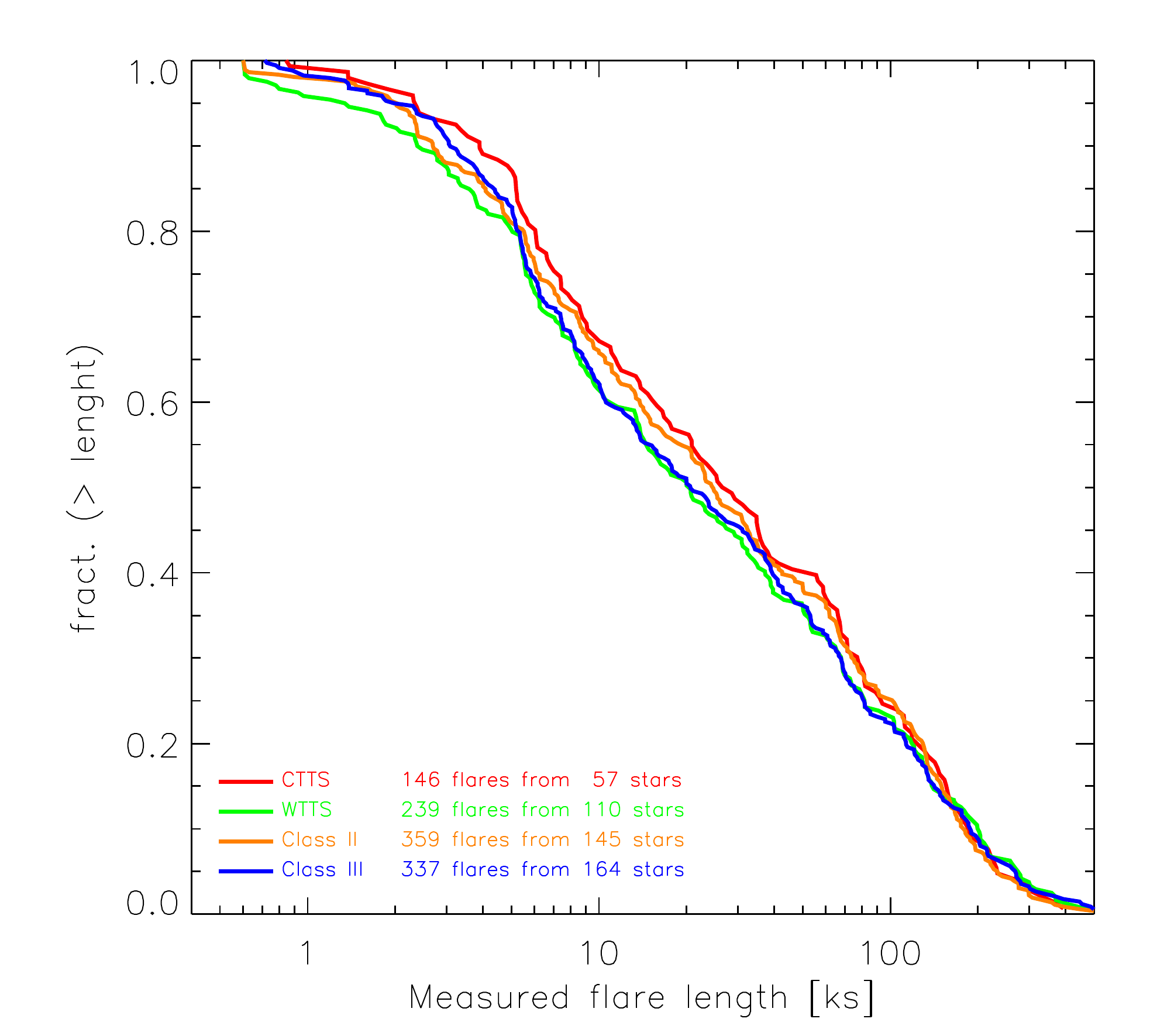}
\caption{[left]: Rate of occurrence of flares as a function of minimum number of flare counts for our four main samples. The legend indicates the color of the lines representing each sample as well as the respective numbers of detected flares and sources. [right]: Cumulative distribution of flare durations for the same four samples.}
\label{fig:flares}
\end{figure*}

To quantify the flaring activity for the same samples of CTTSs, WTTSs,
Class\,II, and Class\,III stars discussed above, we applied the
direct-detection method discussed in \S\,\ref{sect:flare_analysis}.
Figure\,\ref{fig:flares} shows the rate of flares as a function of
minimum flare counts and the cumulative distribution of measured
durations for our four main samples. We conclude that the flares from
the four samples have very similar properties, both in terms of rate of
occurrence and in their duration.

We note, however, that this observation alone does not allow us to
conclude that the contribution of flares to the variability amplitudes,
$VA$s, is the same for all samples. Indeed, $VA$s are a measure of {\em
relative} variability while our flare rates indicate that the {\em
absolute} intensity of flares is the same. However, since we have seen
in \S\,\ref{sect:samples} that CTTSs (and, to a lesser extent, Class\,II
stars) are on average fainter than WTTSs and Class\,III stars, the same
absolute variability (i.e. flare rate) may translate into a more
pronounced relative variability (i.e. $VA$) in CTTSs than in WTTSs, as
is indeed observed. It remains to be seen, however, if this effect is
strong enough to justify quantitatively the observed differences in
$VA$s among our samples.

In the following section we resort to Monte Carlo simulations to
constrain the contribution of flaring activity to the variability
amplitudes as a function of timescale.

\section{Discussion}
\label{sect:discussion}

We now use the results obtained in \S\,\ref{sect:results} to discuss the
origin of variability in the X-ray band. We consider two separate
mechanisms: magnetic flares and rotational modulation. First, we try to
understand to what extent the observed variability and the differences
we observe among subsamples can be explained in terms of flares. We then
focus on rotational modulation and, finally, discuss the physical
implications for the considered samples.

\subsection{The contribution of flares}
\label{sect:disc_flares}

The  effect of flares on the fractional variability amplitudes
depends, in addition to flare intensity and frequency, also on their
durations and on the quiescent or {\em characteristic} count-rate on top
of which they are observed. For example, for a flare of a given intensity
(emitted energy or number of detected counts), a longer duration or a
higher characteristic count-rates will result in lower relative
variability amplitudes. Moreover, the trend of flare-induced $VA$s with
timescale will depend on flare lengths: since flares are expected to
occur stochastically with no particular timescale, rising trends are
only possible on timescales shorter than a few decay times, $\tau$.

We have seen that our four main samples share similar frequencies and durations for detectable flares (\S\,\ref{sect:flare_freq}). However,
since the samples have mean fluxes (or count-rates) that differ by up to
a factor of $\sim2$, (\S\,\ref{sect:samples}), we cannot
straightforwardly conclude that the effect of flares on the $VA$s is the
same. To assess the effect of flares on $VA$s realistically
(even if not rigorously), we ran Monte Carlo simulations of light
curves that attempt to reproduce the following observables: $i)$ the
time-averaged source count-rates, $ii)$ flare frequencies, and $iii)$
flare durations. These simulations are fully described in Appendix
\ref{app:MCflares}, and lead to the conclusion that flares are likely
{\em not} the dominant source of variability as measured by the $VA$s
and cannot explain the differences between our samples. This is
particularly true at long timescales. We indeed show that for
realistic flares durations the run of $VA$s with timescale remains
rather flat.

\subsection{Rotational modulation}
\label{sect:rotmod}

We have seen that the run of $VA$ with phase lag shows signatures of
rotational modulation, both for WTTS/Class\,III stars and, most
strikingly,  for CTTSs in the soft X-ray band. To investigate
this scenario in greater detail we computed the expected $VA$ signal
for a purely sinusoidal modulation with a given relative
amplitude $A$ ($Flux=1.0+A \sin \phi$).

\begin{figure*}[!th!]
\centering
\includegraphics[width=9.0cm,clip]{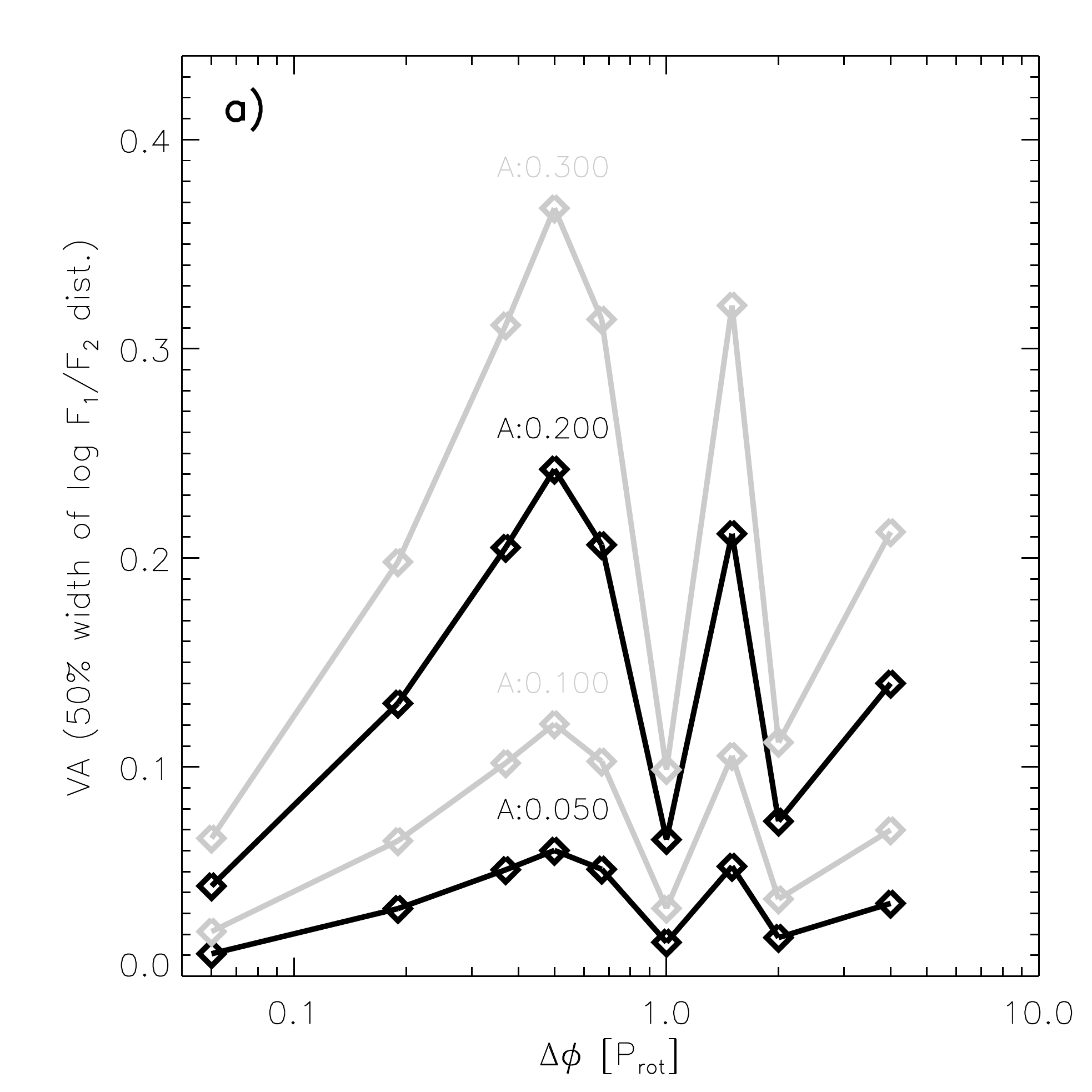}
\includegraphics[width=9.0cm,clip]{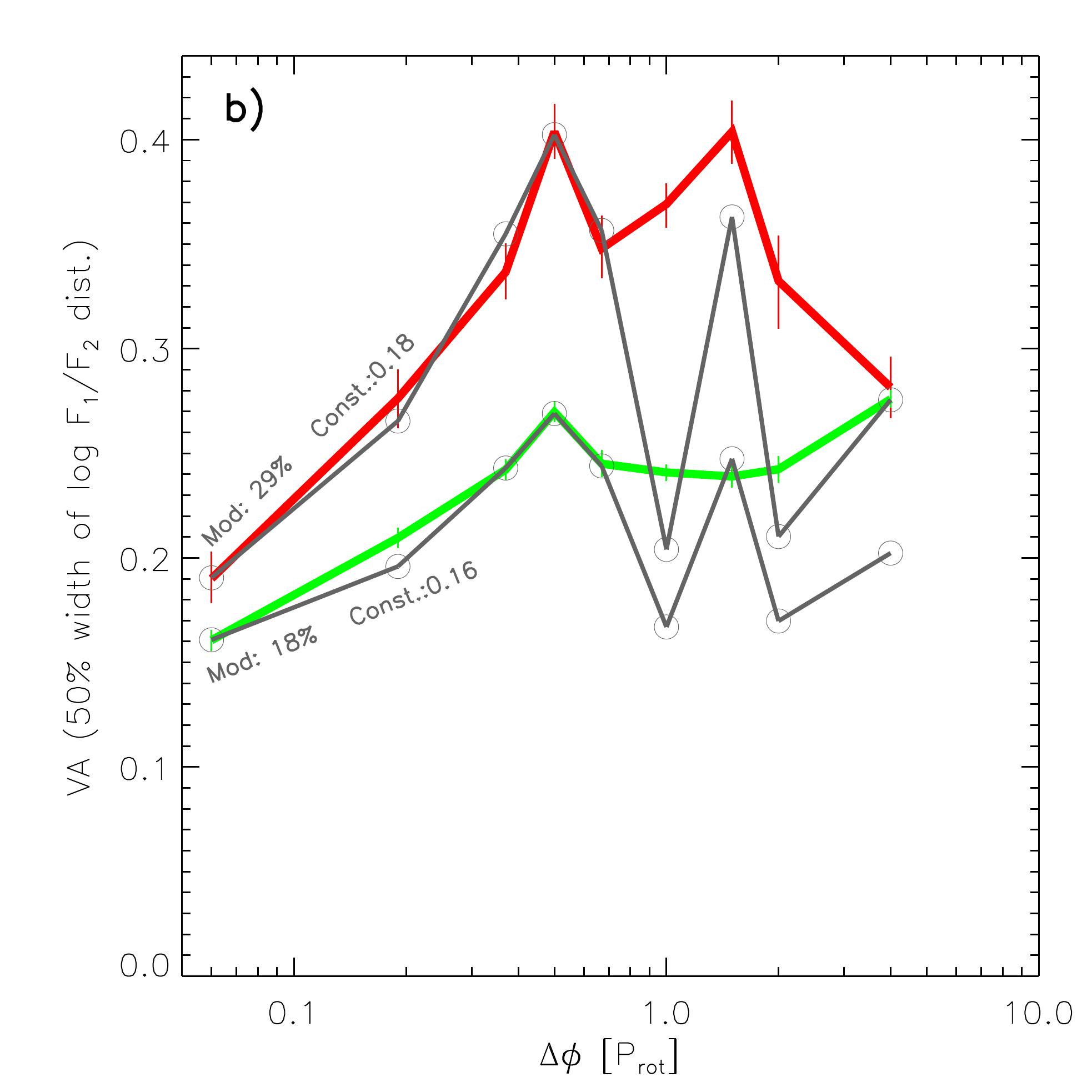}
\caption{[left]: $VA$ vs. rotational phase for simulated lightcurves. The
simulations reproduce sinusoidally varying stars, observed with the COUP
observation and analyzed in the same way as the real data. Four curves show results for
modulation amplitudes from 5\% to 30\% , as labeled. [right]: Comparison of the
trends observed  in the soft band for WTTS and CTTS (green and red lines, respectively) with two
simulated $VA$ curves  (labeled gray lines) obtained by summing in quadrature
the $VA$s for pure sinusoidal modulation (with amplitudes of 18\% and 29\%) with
constant non-modulated $VA$s (0.16 and 0.18\,dex, see labels). Modulation
amplitudes and constant contributions are chosen to reproduce the observed $VA$s
at $\Delta \phi$=0.06 (the leftmost points) and  $\Delta \phi$=0.5 (the first
maxima).}
\label{fig:rotmod_simul}
\end{figure*}

For this simple experiment we drew rotational periods
randomly from either a uniform distribution (between one and ten days) or
from the actual distribution of rotational periods in our samples, with
almost identical results. Rotational phases at the beginning of the
observation were also chosen randomly. Figure\,\ref{fig:rotmod_simul}
[left] shows the $VA$s vs. phase-lag trends expected for relative
amplitudes ranging from 5\% to 30\%. For better comparison with
observations we then added the ``modulated'' $VA$ signal to a constant
$VA$ level, meant to reproduce the contribution of all non-modulated
variability sources such as flaring (see \S\,\ref{sect:disc_flares}), as
well as Poisson noise. Because the modulated and constant
contributions have to be added in quadrature, the absolute value of the
latter {\em smaller} contribution has an important effect at $\Delta
\phi$ values where the modulated $VA$s are low, but little effect on the
peaks at $\Delta \phi$=0.5 and 1.5\,$P_{rot}$.

Figure\,\ref{fig:rotmod_simul} [right] compares the resulting simulated
$VA$s with the observed trends for WTTS and CTTS in the low-energy X-ray
band (lower and upper gray lines).  Modulation amplitudes and constant
contributions were chosen to reproduce the observed $VA$s at $\Delta
\phi$=0.06 (the leftmost points) and  $\Delta \phi$=0.5 (i.e. where the
first maxima are observed). For both WTTSs and CTTSs, very good
agreement with the observed trends can be obtained for $\Delta\phi \le
0.67$\,d, i.e. up to the datapoints following the peaks at 0.5\,d, by
adding in quadrature a constant contributions of $\sim$0.16\,dex and
$\sim$0.18\,dex and for modulation amplitudes  of 18\%  and $\sim$29\%,
respectively. For higher values of $\Delta \phi$ the agreement is not as
good: for CTTSs, even though the maximum predicted at $\Delta
\phi$=1.5\,$P_{rot}$ is indeed observed, the minimum predicted at
$\Delta \phi$=1.0\,$P_{rot}$ is too deep and the one at 2.0\,$P_{rot}$
is not observed. Similar considerations apply to WTTSs. The quantitative
disagreement may indicate that the actual variability is not sinusoidal,
as assumed, and/or, maybe more importantly, that the X-ray modulation
periods are not the same as the rotational periods (see below), and/or
that the structures responsible for the modulation (e.g. coronal active
regions or absorbing circumstellar structures) evolve on timescales of
a few rotational periods so that the modulation signal is lost for $\Delta
\phi >$1-2\,$P_{rot}$. Given the above considerations, we conclude that
the clearer modulation signal observed for CTTSs in the soft X-ray band
indeed suggests stronger modulation than for WTTSs.

A similar analysis for the hard X-ray band, for which only WTTS show
clear indications of rotational modulation, results in a best estimate
for the modulation amplitude of WTTS of 24\% and a non-modulated
contribution to the $VA$s of $\sim$0.22\,dex. For CTTS, instead, our
simple model is not able to account for the observed trend, indicating
that rotational modulation plays a relatively less important role with
respect to other sources of variability such as flares (which cannot
account for the increase of $VA$s with timescale, however) and the
evolution of emitting or absorbing structures (i.e. active regions and
circumstellar matter).

Our detection of rotational modulation in the soft band is fairly
consistent with the conclusions of \citet{fla05}, who found signatures of
rotational modulation in 23 out of the 233 COUP sources in their parent
sample, with X-ray amplitudes in the {\em full energy band} ranging from
20\% to 70\%.  No evidence was found of a correlation between accretion
and the detection of X-ray modulation and/or modulation amplitudes,
apparently at odds with our findings that suggest larger modulation
amplitudes for CTTSs, at least in the soft X-ray band. The small sample
size of \citet{fla05}, differences in the samples and the fact that only
the full X-ray band was considered might very well explain the potential
discrepancy.

Our samples significantly overlap with that of \citet{fla05}, who
adopted for their analysis a lower count-rate threshold with respect to
ours and excluded stars with short rotational periods, $\rm
P_{rot}<$2\,d. In our sample of CTTSs (WTTSs), for example, 17 (51) out
of the 19 (72) stars in our ``full band'' sample are in common between
the two studies. Among these, \citet{fla05} claimed the detection of
rotational modulation in three CTTSs and ten WTTSs.

Although \citet{fla05} did not detect a difference between CTTS and WTTS with
regard to the incidence of X-ray rotational modulation or modulation
amplitudes, a hint was  found at the 1\% significance level that CTTSs
(and Class\,II stars) have X-ray modulation periods that are
preferentially one half of the optically determined periods. The same
hint is present in the subsample in common with the present study, since
all three CTTS for which a periodicity was detected by \citet{fla05} were found
to have $\rm P_{X-ray}\sim0.5\,P_{opt}$, while the same is true for
only two out of ten WTTSs. If, as it seems to be the case for CTTSs,
X-ray periods are shorter than the optically determined period in a
significant fraction of cases, our interpretation of the peaks in the
runs of $VA$ vs. phase lag (these latter computed using $\rm
P_{opt}$) will be affected. In the cases for which $\rm
P_{X-ray}\sim0.5\,P_{opt}$ we would indeed expect $VA$ peaks at
0.25\,$\rm P_{opt}$,  0.75\,$\rm P_{opt}$, etc. Although these peaks
would not be detectable with the $\Delta\phi$ binning we adopted in
our analysis (e.g. Fig.\,\ref{fig:VarAmp}), they might fill in the
minima between the maxima expected for the $\rm P_{X-ray}\sim P_{opt}$
case (e.g. between the 0.5\,$\rm P_{opt}$ and 1.5\,$\rm P_{opt}$). As
already mentioned, this might explain why the observed peaks
and minima, e.g.,  in the $VA$s for CTTSs in the soft X-ray band, are
not as pronounced as predicted by our simulations of sinusoidally
modulated lightcurves (Fig.\,\ref{fig:rotmod_simul}).

\subsection{Implications}
\label{sect:discPhyImp}

Building on the above discussion, we now speculate on the
physical mechanisms responsible for the observed X-ray variability, with
particular reference to the difference between CTTSs and WTTSs, i.e. by
our definition stars undergoing ``strong''  and ``no'' accretion,
respectively. We focus on the following results: $i)$ flares are a
certain source of variability in all our subsamples; $ii)$ most
of the variability, however, cannot be explained by flares of the
observed duration, especially at long timescales; $iii)$ the higher
variability of CTTS/Class\,II stars with respect to WTTS/Class\,III
stars cannot be explained by different flaring activities; $iv)$ flux
modulation due to stellar rotation is a good candidate to explain at
least part of the non-flaring variability; $v)$ in the soft X-ray band,
in particular, CTTSs show significant signatures of rotational
modulation, suggesting higher modulation amplitudes than in WTTSs; $vi)$
no indication of rotational modulation is observed for CTTSs in the
harder X-ray band.

\subsubsection{The coronae of CTTSs}

The enhanced rotational modulation observed for CTTSs in the soft X-ray
band with respect to the hard band may indicate that the azimuthal
distribution of {\em observable} structures emitting {\em soft} X-rays
is more inhomogeneous than that of those responsible for the hard
X-rays.  Two scenarios are qualitatively consistent with this
conclusion: $i)$ modulation of the X-ray emission from accretion spot(s)
or from an accretion-fed corona \citep[as proposed by, e.g.][]{bri10},
both expected to be predominantly soft, and $ii)$ modulation of the
line-of-sight extinction toward a significant fraction of the coronal
structures, again affecting mostly soft X-rays. The first accretion
hypothesis builds on the observations of X-ray emission from accretion
spots, including recent observations of its rotational modulation
\citep[e.g.][]{arg11,arg12}. The second absorption hypothesis could
be related to the periodic obscuration of large parts of the stellar
surface by dusty structures in the inner disk, or by the gaseous
accretion streams in co-rotation with the star. Obscuration of the
photospheric emission by disk warps has been extensively observed for AA Tau \citep{bou99,bou03,men03,bou07,gro07} and has been inferred to be
common in CTTSs by \citet{ale10}. Absorption by accretion streams was
predicted to be effective in the X-ray regime by \citet{gre07}. The same
accretion and absorption hypotheses were recently also proposed
by \citet{fla10} to explain the correlation between the X-ray and
optical variability in the NGC\,2264 star-forming region, found only for
CTTSs and in the {\em soft} X-ray band. \citet{fla10} favored the
absorption/AA Tau-like scenario mainly because the contribution of
the accretion shock emission to both the optical and the soft X-ray band
was estimated to be too low to explain the large observed variability
amplitudes.

The same argument applies here to the X-ray emission from accretion spots: in all cases this emission has been revealed by the observation of emission lines from cool and dense plasma,  detected by the {\em Chandra} and {\em XMM-Newton} high-resolution spectrographs. Even for the most extreme case known, TW\,Hya,  the accretion shock emission is too soft and too faint to contribute a large part of the broad-band X-ray flux detected by imaging instruments, such as the {\em Chandra} ACIS \citep{dup12}. 

Moreover, the presence of such a significant soft contribution would be evident in the observed CCD spectra and would {\em require}, for CTTSs only, a cool thermal component in the X-ray spectral fits. The AA Tau-like hypothesis, i.e. the heavy absorption of {\em part} of the emission from the gaseous accretion streams, {\em might} instead find support from the time-average CCD spectra if: $i)$ two thermal components of similar temperaure, but subject to very different absorptions, were required to fit the spectra or, $ii)$ the absorptions derived from the fits to the X-ray spectra were significantly higher than those derived from the optical colors and spectral types. The former scenario would indicate the presence of two {\em spatially distinct} plasma components of comparable emission measure, only one of which intercepts the additional absorber (disk warp or accretion stream) along the line of sight, while the emission of the other is only subject to interstellar absorption. The latter scenario might instead result if the circumstellar structures absorb the great majority of the emission.

Unfortunately, none of the above predictions is borne out by the CCD
spectra of the 11 (19) CTTSs and the 62 (72) WTTSs entering in our $VA$
analysis in the soft (full) X-ray band\footnote{As for the presence of a
significant accretion-spot contribution, we compared the temperatures of
the cool thermal components needed to fit the CCD spectra of CTTS and
WTTS, as reported by \citet{get05}. We found that CTTSs host, if
anything, systematically hotter plasma than WTTS, at odds with the
accretion hypotesis. As for the possible evidence of AA Tau-like
behavior within our low-energy sample, we found no evidence of double
absorptions, both as indicated by \citet{get05}, who flagged spectral
fits with evidence of this situation, and upon new examination of the
spectral fits. We also found no suggestion that CTTS have a peculiar
$N_H$-$A_V$ relation with respect to  WTTS.}. However, we argue that
while for the accretion scenario we should indeed expect a detectable
soft emission component, if it were responsible for the observed
rotational modulation, the non-detection of an absorption ``anomaly'' in
CTTS is actually consistent with the absorption/AA Tau-like
scenario. We reach this conclusion through Monte Carlo simulations that
aimed to reproduce the observation of X-ray spectra composed of two
intrinsically similar thermal components absorbed by different gas
columns. These simulations, described in detail in
Appendix\,\ref{app:spMC}, show that the detection of the heavily
absorbed component is extremely difficult, if not impossible, using
current-quality data. We therefore conclude that the CCD spectra
disfavor the accretion scenario as an explanation of the observed X-ray
variability, but are compatible with the absorption/AA Tau-like
scenario.

\subsubsection{Time-averaged activity levels of CTTS and WTTS}

The absorption scenario is appealing also because it may provide an
explanation not only for the X-ray variability, which is the main subjet
of the present study, but also for the previously recognized differences
in the time-averaged X-ray activity levels of CTTSs  ($L_X$ or
$L_X/L_{bol}$)  when compared to WTTSs:  their on-average lower and more
scattered activity levels at any given mass, or $L_{bol}$.

Obscuration of large parts of the coronal plasma by the dense
accreting columns or disk warps may indeed explain the observation that
CTTSs have, on average, 2-3 times lower activity levels than WTTSs, even
assuming that their coronae are similar. As shown in Appendix
\ref{app:spMC}, the obscured plasma cannot be detected and
accounted for with current-quality data. Moreover, for sufficiently high
absorbing column densities, $N_H\gtrsim 3\times 10^{22}$\,cm$^{-2}$, and
sufficiently extended absorbing structures, intercepting $>$1/2 of the
coronal plasma, the retrieved X-ray luminosities will indeed be less
than one-half of the full coronal luminosity, with no discernible effect
on the $N_H$ estimated from the spectral fits. The observed magnitude of
our $VA$s, up to $\sim$0.5\,dex for timescales of $\sim$10 days and
maybe larger for longer timescales, if attributed to periodic
obscurations of parts of a corona from circumstellar structures, imply
that the obscured fraction is significant.

The estimate for the modulation amplitudes obtained in
\S\,\ref{sect:rotmod} for the soft X-ray band, $\sim$30\%, if attributed
to absorption implies a decrease in average flux on the same order.
Therefore, to explain a reduction of a factor of $\sim$2 in the
observed emission of CTTSs, on average, we would need to assume that an
additional fraction, $\sim$20\%, of the emission is permanently
obscured. This seems reasonable.

We now consider the difference in the scatter of the observed activity
levels, making the reasonable assumption that the time variability of
the stars in our samples is uncorrelated. Regardless of its physical
origin, time variability {\em on long timescales} will therefore
contribute to the observed scatter in activity levels at any given mass
or $L_{bol}$. Our finding that CTTSs are more variable with respect to
WTTSs may therefore explain the observed differences at least
qualitatively.

We now aim to assess the contribution of variability to this scatter
more quantitatively. Indeed, our measure of variability, the $VA$s,
would translate directly into a contribution to the 50\%
scatter\footnote{The difference between the 25\% and the 75\% quantiles
of the residuals.} of activity levels (strictly speaking averaged over
30\,ks) if they could be measured on the longest timescales (with the
reasonable assumption that they level off at some timescale). Since we
only measured variability amplitudes for timescales up to $\sim$10 days,
we consider our maximum $VA$ for a given sample, $VA(10\rm d)$, as a
lower limit to the long-term variability, $VA(\infty)$.

$VA(10\rm d)$ is 0.47\,dex for CTTS and 0.34\,dex for WTTSs (cf. the
upper-left panel of Fig.\,\ref{fig:VarAmp}). We may compare these
numbers with the scatter in the {\em characteristic} $L_X$ vs. $L_{bol}$
relation found by \citet{pre05a} for larger samples of CTTS and WTTS in
the COUP dataset, where the definition of WTTS and CTTS is the same as
adopted here. \citet{pre05a}, however, only  report the 1$\sigma$
dispersions (0.72\,dex and 0.52\,dex for CTTS and WTTS, respectively).
To compute 50\% dispersions for direct comparisons with our $VA$s, we
repeated the $L_X$ vs. $L_{bol}$ regression analysis with the COUP
sources using three sets of X-ray luminosities: the {\em characteristic}
values, $L_{X,char}$, the time averages $L_{X,av}$ (over the 850\,ks
COUP exposure), and those relative to the individual 30\,ks segments we
employed for our variability analysis, $L_{X,30ks}$. These latter values
should allow the most direct comparison between $VA$s and scatter, and
were derived by scaling the time-averaged $L_X$ to the ratio between the
count-rate in each 30\,ks segment and the average count-rate. 
Throughout the regression analysis we used maximum likelihood ($\chi^2$)
linear fits that, by definition, yield minimum scatter.

\begin{figure*}[]
\centering
\includegraphics[width=8.0cm]{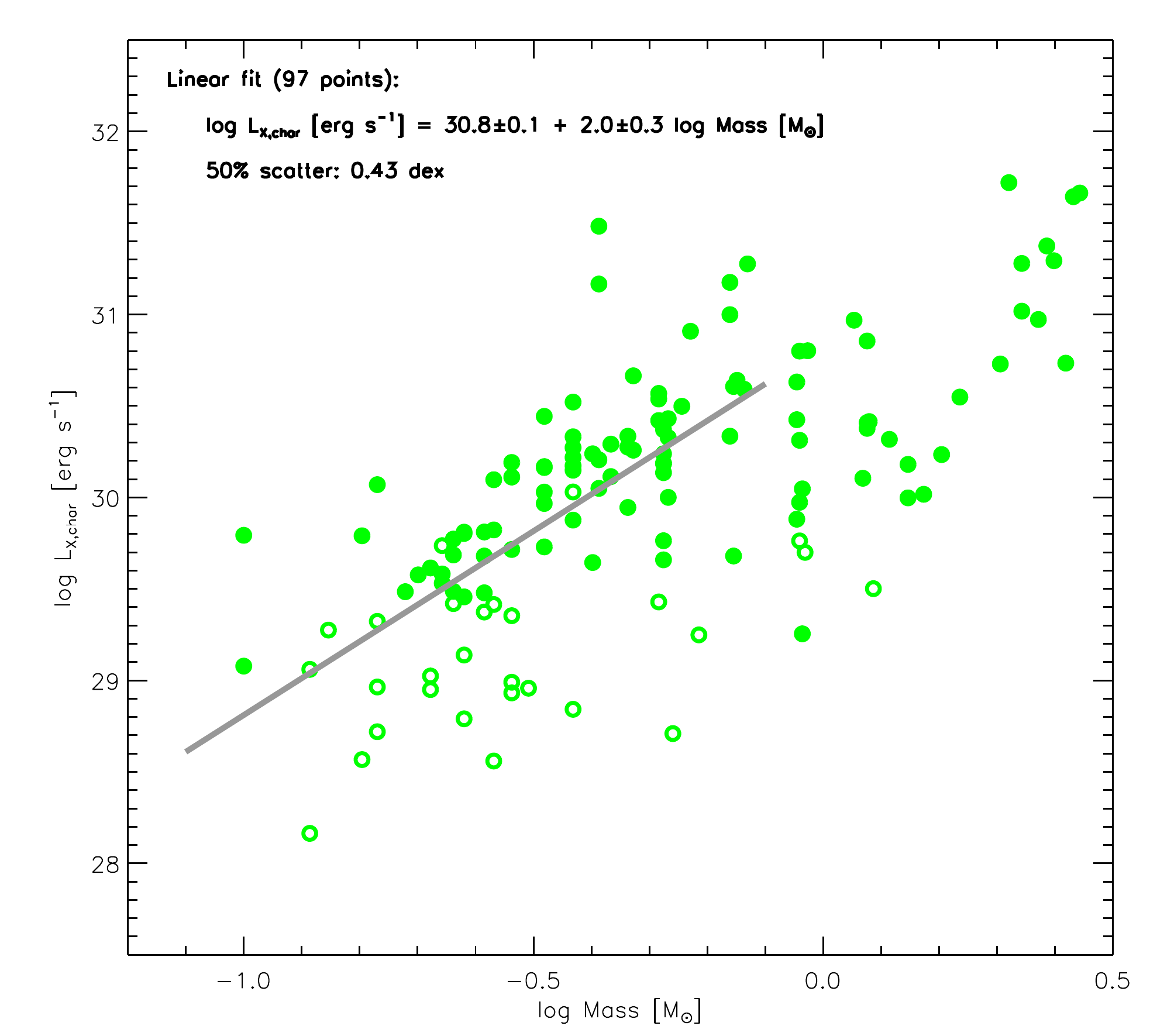}
\includegraphics[width=8.0cm]{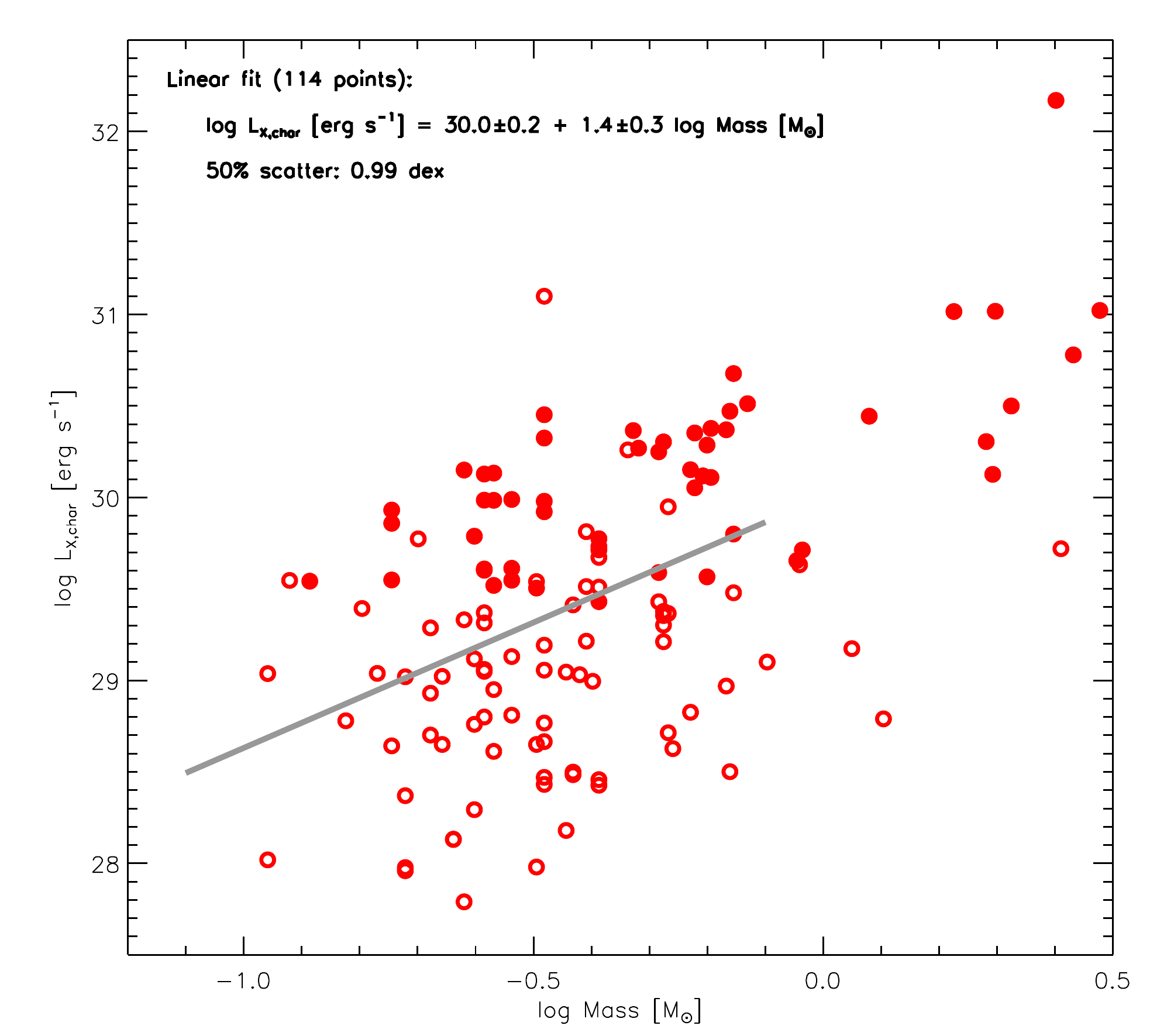}
\caption{{\em Characteristic} $L_X$ vs. stellar mass for WTTSs and CTTSs (left and right
panels, respectively). The samples are selected in the same way as for our variability
study except that in this case we did not impose an X-ray flux (or 
count-rate) limit. Filled circles refer to the count-rate-limited sample 
used for the variability study. Linear fits are computed from all points 
(both filled and empty symbols) within the x-axis range spanned 
by the gray lines that represent them.
}
\label{fig:act_ml}
\end{figure*}

Our regression analysis for the $L_{X,char}$ vs. $L_{bol}$ relation
of WTTS and CTTS yields very similar results to those presented by
\citet{pre05a} (c.f. their Fig.\,17). Best-fit parameters and 50\%
dispersions for these regressions and for those computed using
$L_{X,av}$ and $L_{X,30ks}$ are listed in Table\,\ref{tab:scatters}.
The same table also shows the results of regression analyses of the
trends of $L_X$ with mass for stars with masses below 0.8\,M$_\odot$,
i.e. for fully radiative stars, as shown in Fig.\,\ref{fig:act_ml} (for
$L_{X,char}$). Since the dispersions around the maximum likelihood
relations are systematically lower (by 0.07-0.15\,dex) than for the
relation with $L_{bol}$, we took these lower values as the measure of
the true scatter of activity levels.

\begin{table}[] 
\caption{Parameters of best-fit linear regressions between $L_X$ (three different sets of values, see text) and $L_{bol}$ or mass (for M$<$0.8\,M$_\odot$), and resulting 50\% scatter.} 
\label{tab:scatters}
\centering 
\scriptsize
\begin{tabular}{ll|rr|rr|rr} 
\hline\hline
 \multirow{2}{*}{Y} & \multirow{2}{*}{X} & \multicolumn{2}{|c}{a [erg s$^{-1}$]} & \multicolumn{2}{|c}{b} & \multicolumn{2}{|c}{50\% scatter [dex]} \\
&  & CTTS & WTTS  & CTTS & WTTS  & CTTS & WTTS\\
\hline
$L_{X,30\,ks}$  	& $L_{bol}$ 	&  29.4 & 30.1 & 0.5  & 1.1 & 1.18  &  0.65 \\
$L_{X,av}$				& $L_{bol}$ 	&  29.7 & 30.3 & 0.4  & 1.0 & 1.18  &  0.63  \\
$L_{X,char}$ 			& $L_{bol}$ 	&  29.5 & 30.1 & 0.4  & 1.0 & 1.06  &  0.58 \\
\hline
$L_{X,30\,ks}$ 		& $Mass$   	&  30.0 & 30.9 & 1.4  & 2.2 & 1.07  &  0.57 \\
$L_{X,av}$				& $Mass$  	&  30.2 & 31.0 & 1.4  & 2.0 & 1.04  &  0.53 \\
$L_{X,char}$ 			& $Mass$  	&  30.0 & 30.8 & 1.7  & 2.0 & 0.99  &  0.43 \\
\hline   	
\end{tabular} 
\end{table}

When comparing these observed scatters in $L_X$ with our $VA$s, we must
acknowledge that the $VA$s are computed from smaller (and
biased) subsamples with respect to the scatters. Fig.\,\ref{fig:act_ml}
shows with filled and empty symbols the stars belonging
and not belonging to the count-rate limited subsample used for $VA$s
($>1$count/Ks, or, roughly, $\log L_X>29.5$\,erg\,s\,$^{-1}$). In the
following we assume that the $VA$s of X-ray-faint
stars, which also have statistically lower $L_{bol}$ and mass, are
similar to those of their brighter counterparts.

All dispersions listed in the last two columns of
Table\,\ref{tab:scatters} for CTTSs and WTTSs, and in particular those
for $L_{X,30ks}$ which should be more directly comparable to the $VA$s,
are higher than the observed $VA$s(10\,d): 0.34\,dex for WTTS and
0.47\,dex for CTTSs. In performing this comparison, however, we should
consider two facts that might explain this differenece: first, as
mentioned above, our $VA$s are computed for subsamples that, compared to the full samples, are biased toward high masses and
bolometric luminosities. An extrapolation of our results on the
dependence of $VA$s on mass (Fig.\,\ref{fig:VarAmp_mass}) suggests that
low-mass stars not included in our variability study may have higher
$VA$s, and indeed Fig.\,\ref{fig:act_ml} may suggest an increase of the
scatter at low masses. Therefore, our $VA$s might be underestimated if
applied to the whole sample down to the lowest masses and $L_{bol}$.
Second, and maybe most importantly,  part of the observed scatter in the
correlations between $L_X$ and $L_{bol}$ or mass will be due to
statistical uncertainties in both quantities, e.g. introduced by the
correction of the observed X-ray flux for the individual stellar
absorptions, and/or by the derivation of stellar masses through
placement of the star in the theoretical HR diagram and comparison with
PMS evolutionary models, and/or by the inclusion in the samples of
unresolved binaries. Following the discussion in \citet{pre05a}, we
estimate that the 50\% scatter introduced by the above uncertainties on
the $L_X$ vs. mass relation is $\sim$0.43\,dex\footnote{Derived summing
in quadrature, 0.2\,dex (uncertainties on masses), 0.15\,dex ($L_X$),
and 0.15\,dex (unresolved binaries), and converting this 1$\sigma$
scatter to a 50\% scatter assuming that the distributions are
Gaussian.}. This contribution is to be subtracted (in quadrature) from
the observed scatters before comparison with $VA$(10\,d).

Given these considerations, we thus argue that for WTTSs, a significant
part of the scatter in $L_X$ at a given mass might indeed be justified
by variability on the timescale of $\sim$10 days\footnote{This is also
confirmed by the scatter measured in the $L_{X,char}$ vs. mass relation,
i.e. using the X-ray luminosity averaged over 10\,d, and after excluding
flares, which is on the same order as the estimated measurement
uncertainties.}. The scatter for CTTSs is, however, significantly larger
than the corresponding $VA$, even considering uncertainties, and cannot
be explained in this way. Variability on longer timescales might,
however, be responsible for most of the scatter in this case. A
forthcoming study of longer-term variability will aim to address this
point. In the absorption (or AA Tau-like) scenario, which we invoked to
explain the evidence of significant rotational modulation in the soft
X-ray band, this long-term variability might be naturally attributed to
the year-over-year evolution of circumstellar structures (disk warps and
accretion streams), and the ensuing variability in the amplitude of the
shading of coronal structrures (both modulated and unmodulated). This
kind of evolution is, indeed, well documented for AA Tau, and may also
be inferred from the observed evolution of the structure of the magnetic
field in V2129\,Oph \citep{don10}.

\section{Summary and conclusions}
\label{sect:summconc}

We have performed a study of X-ray variability among the young stars in
the ONC, as observed with the 850\,ks long COUP observation. By
investigating  variability for different stellar samples as a function
of timescale and rotational phase (for stars with known rotational
period), we have found a number of novel results:

- CTTSs are significantly more variable than WTTSs at all timescales
and in all X-ray spectral bands.

- For all samples variability amplitudes increase with increasing timescale at least up to $\sim 10$ days, i.e. the longest we probed.

- Signs of rotational modulation are visible, most clearly for CTTSs and
in the soft 500-1500\,keV band.

- Among low-mass Class\,II and Class\,III stars, variability decreases
with increasing stellar mass.

We speculate that the difference in variability between CTTSs and WTTSs
may be explained assuming that the X-ray emission of CTTS is affected by
time-variable absorption caused by circumstellar structures, such as
warps in the inner disk and/or accretion streams, in co-rotation, or
close to co-rotation, with the star. This suggestion is appealing
because, assuming that the coronae of CTTSs and WTTSs are similar, it
may also explain why CTTSs have lower and more scattered X-ray emission
levels than WTTSs.

\begin{acknowledgements}

We thank Jeremy Drake for stimulating discussions and careful
reading of the manuscript.

\end{acknowledgements}

\bibliographystyle{aa} 
\bibliography{bibtex.bib}

\appendix

%
%
%
%

\section{Monte Carlo flare simulations}
\label{app:MCflares}

As discussed qualitatively in the main text
(\S\,\ref{sect:disc_flares}), the effect of flares on the fractional
variability amplitudes, the $VA$s, depends on the intensity, frequency,
and duration of flares, as well as on the quiescent or {\em
characteristic} count-rate on top of which they are observed. In this
appendix we employ Monte Carlo simulations to assess,
quantitatively and realistically, the contribution of flaring activity
to the $VA$s. We are particularly interested in determining whether
flares can explain the  observed rising trends of $VA$s with timescale
and the differences among our four main samples (CTTSs, WTTSs,
Class\,II, and Class\,III stars) .

Although these samples share similar frequencies and durations
(\S\,\ref{sect:flare_freq}) for detectable flares, our flare detection
algorithm can only identify flares that individually contribute a
statistically significant signal to the lightcurves. The $VA$s may, however,
be affected also by smaller flares, either individually or by their
superpositions. To assess the effect of flares on $VA$s
realistically, we ran simulations of light curves that attempt to
reproduce the following observables: $i)$ the time-averaged source
count-rates, $ii)$ flare frequencies,  and $iii)$  flare durations. More
specifically, following the results of \citet{car07}, we assumed
that the X-ray emission of our sources is produced by a superposition of
flares with impulsive rise phases and exponential decays, with a given
fixed exponent $\tau$ (at least initially). This last assumption is an
approximation, since the observed flares clearly show a range of decay
times, and was eventually relaxed.

Flares are assumed to occur at an energy-dependent rate, $dR_{fl}(E)$,
with a power law dependence: $dR_{fl}(C)/dC \propto C^{-\alpha}$, where
we substituted the energy with the number of detected flare counts,
$C$. \citet{car07} found $\alpha\sim-2.2$ for low-mass ONC members. We
adopted the same value here, which is well-compatible with the bright
tail of the flare frequency distribution in Fig.\,\ref{fig:flares} (the
turn-off at $\log C\lesssim3$ may be explained by the inefficiency of
the flare-detection process). \citet{car07} was able to reasonably
reproduce the mean characteristics of observed light-curves, as well as
the flare frequency distribution, by assuming that the whole stellar
X-ray emission is caused by flares, with a source-dependent minimum flare
amplitude, $C_{min}$, and total flare frequency, $R_{fl}$$\equiv \int
dR_{fl}(C)$. Note that, if $\alpha<-2$, a given set of $\alpha$,
$C_{min}$ and $R_{fl}$, directly determines the mean source count-rate,
as well as the {\em characteristic} one.

Here, we followed a slightly simplified approach with respect to
\citet{car07}: instead of  trying to reproduce the characteristics of
each individual light curve, i.e. by choosing individual $C_{min}$ and
$R_{fl}$ values to fit the mean and {\em characteristic} fluxes, we
represented all the stars in each of our samples with a single set of
$C_{min}$ and $R_{fl}$ values and thus tried to reproduce the median
count-rate of the stars in the sample.  Moreover, in the last two of the
four simulation sets we discuss below, we also allowed for the presence
of a quiescent non-flaring emission component to better reproduce the
observations, and, in the last set, for a range of flare decay
timescales, $\tau$.

All our simulations were run with $\alpha=-2.2$ and $\alpha=-2.4$ but we
only discuss the case of $\alpha=-2.2$ since the conclusions are
the same for the two cases. We chose $C_{min}$ and $R_{fl}$ so that
$i)$ the expected frequency of simulated flares with $\log C>3$ matches
the observed value in Fig.\,\ref{fig:flares}, $\sim$0.4 flares/Msec for
all samples, and $ii)$ the simulated lightcurves match the median
count-rate of our samples.  In particular we ran simulations to
reproduce median count-rates of 3, 4, and 6\,cnts/Ksec (cf.
Fig.\,\ref{fig:crate_dist}). 

For each simulation set, i.e. for each set of parameters, $\alpha$,
$C_{min}$, $R_{fl}$, and $\tau$, we generated 1000 simulated
lightcurves, covering the same temporal observing windows as in the COUP
observation (see e.g. Fig\,\ref{fig:exLC}), and applied the exact same
analysis as previously described for the real COUP sources in our samples,
i.e. the $VA$ analysis as a function of timescale and flare detection.

\begin{figure*}[!th!]
\centering
\includegraphics[width=18.0cm,clip]{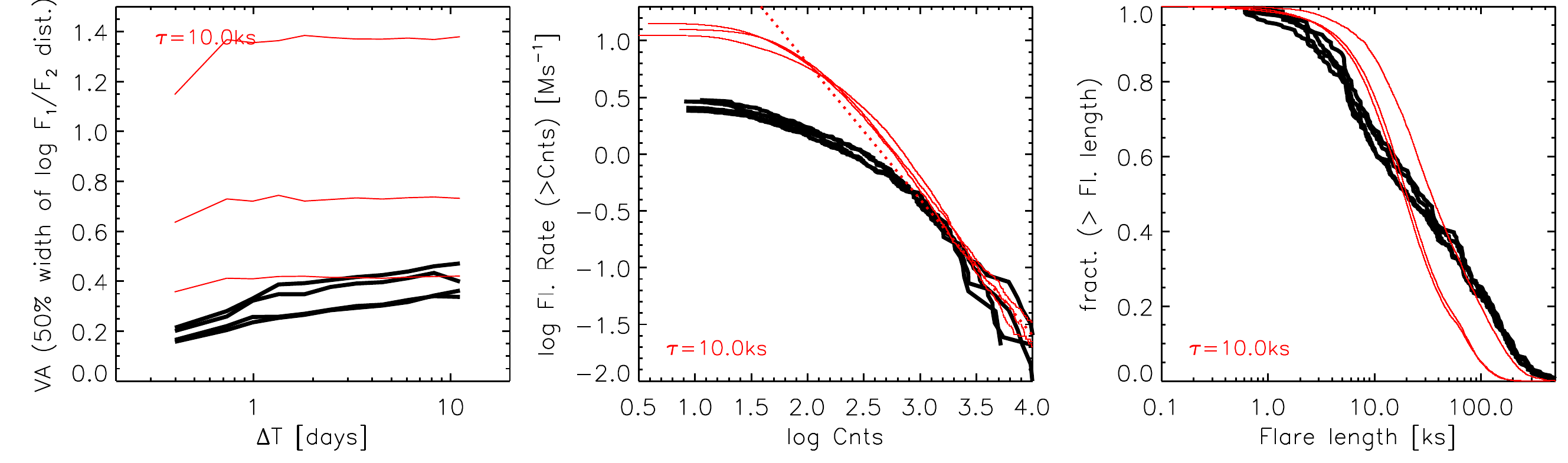}
\includegraphics[width=18.0cm,clip]{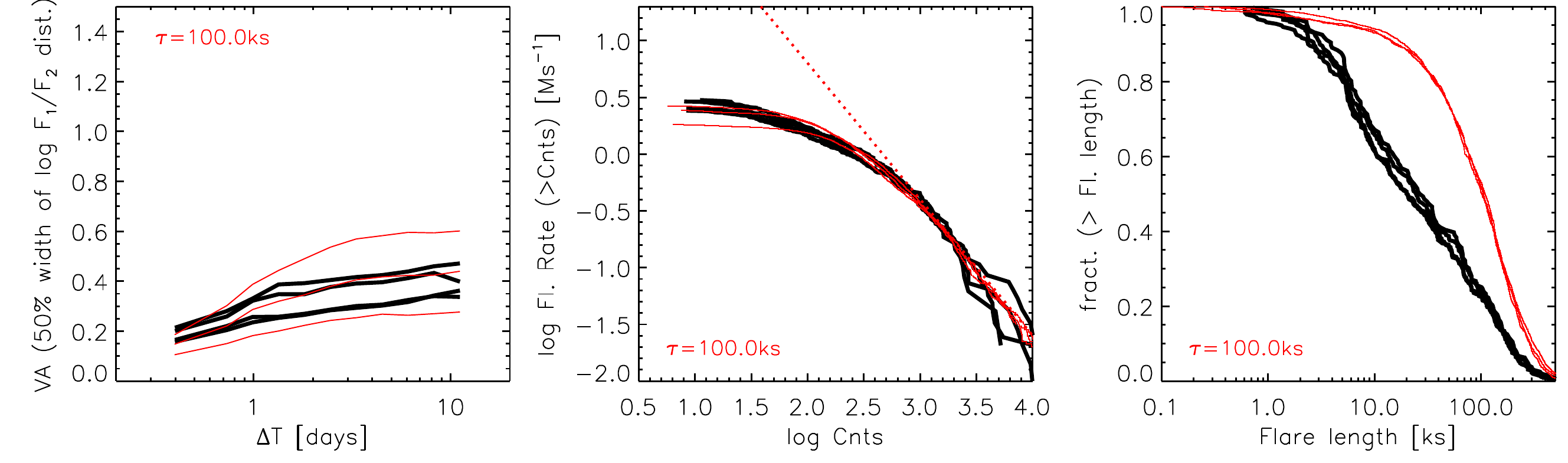}
\includegraphics[width=18.0cm,clip]{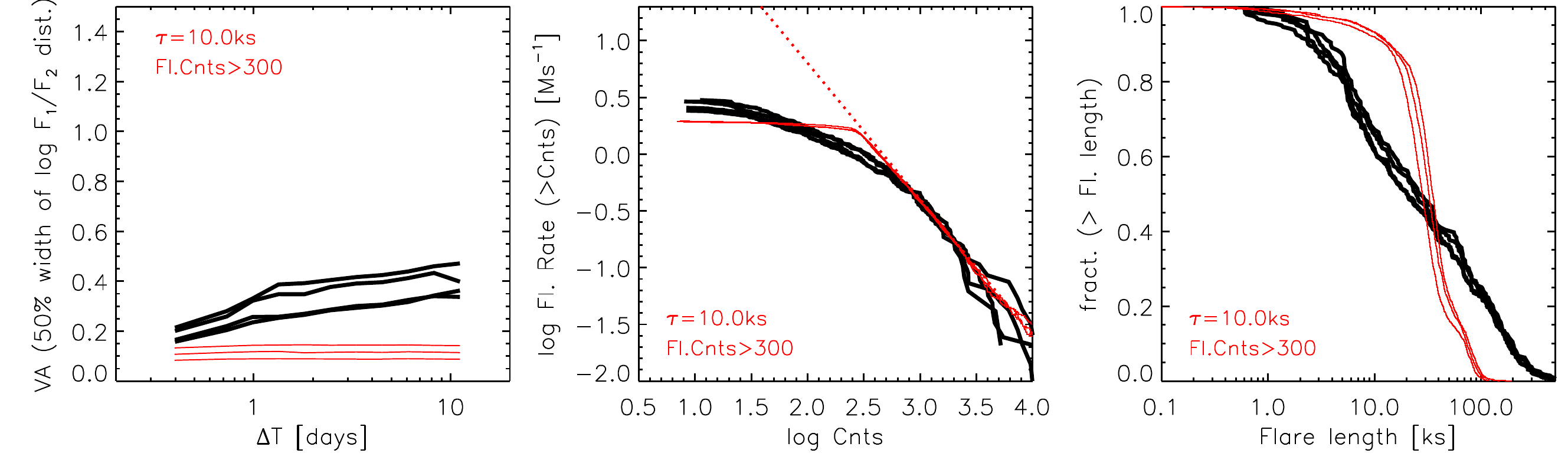}
\includegraphics[width=18.0cm,clip]{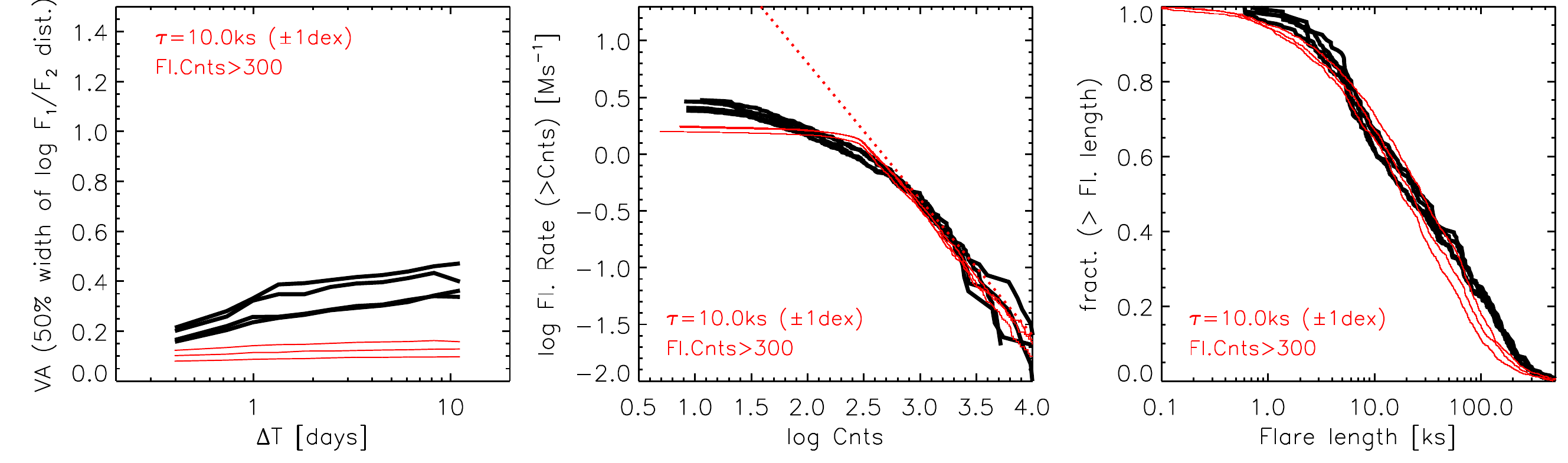}
\caption{Results of our $VA$ and flare analysis on simulated lightcurves
consisting of a superposition of flaring and continuous emission (see
text). From left to right we show  measured $VA$s, flare frequencies,
and flare durations. In all cases the three red lines refer to
simulation sets that reproduce a given median count-rate: 3, 4, and
6\,cnts/Ksec, spanning the median count rates of our four main stellar
samples. Observed quantities for these four samples are drawn in
black, reproduced from Figs.\,\ref{fig:FxRatioDist} and \ref{fig:flares}.
The four rows refer to four different simulation sets that differ
for the assumed flare decay-times and, in last two cases, for a lower cutoff
in flare intensities (in counts), with a corresponding addition of a
continuous emission level. From top to bottom: $\tau$=10\,ks (no
cutoff/continuum), $\tau$=100\,ks (no cutoff/continuum), 
$\tau$=10\,ks  ($>$300 counts plus continuum), and 
$\tau$=10\,ks$\pm1$\,dex ($>$ 300 counts plus continuum) .}
\label{fig:flare_simul}
\end{figure*}

For our first set of simulations we set $\tau$=10\,ks, i.e. similar to
that of {\em most} observed flares. The top  row of
Fig.\,\ref{fig:flare_simul} shows the results. From left to right: the
run of $VA$s vs. timescale, the flare frequency $R_{fl}$ vs. flare
counts, and the distribution of flare durations. In all cases the black
lines show the actual observed functions/distributions for our four
samples, repeated from Figs.\,\ref{fig:FxRatioDist} and \ref{fig:flares}.
The three thin red lines instead show the results of the simulations,
one for each of the median count-rates, 3, 4, and 6\,cnts/Ksec,  that
encompass the observed ones\footnote{Note that in the plot of $VA$ vs.
timescale, higher $VA$s, a measure of relative variability, are
associated with lower count-rates.}. We conclude that our models, while
reproducing the observed flare durations reasonably well, produce light
curves that are much more variable and with many more faint flares than
observed. A possible solution to reduce both $VA$s and the frequencies
of  {\em detected} faint flares is to increase $\tau$. The second row
in Fig.\,\ref{fig:flare_simul} shows that for $\tau$=100\,ksec we indeed
obtain a good match with the observed $VA$s and flare frequencies. For
$VA$s, we even qualitatively match the observed rising trends  with
timescale. However, the simulations do not match the measured flare
durations, a discrepancy that is clearly supported by simple inspection of
the observed and simulated lightcurves. 

In our third simulation set we therefore decided to revert $\tau$ to
10\,ksec and, to lower $VA$s and flare frequencies, to
substitute the flux emitted by small flares below a given counts threshold,
with a quiescent emission. The third row in Fig.\,\ref{fig:flare_simul}
shows the result of setting the minium flare-counts threshold to 300. We
note that $i)$ flare frequencies are now reasonably close to the
observations, especially considering that our sharp cut of flare counts
at 300 is obviously unrealistic; $ii)$ $VA$s are significantly smaller
than observed, especially for long timescales, which leaves room for
other sources of variability such as rotational modulation (of which we
found evidence in \S\,\ref{sect:results_VA_RP}); $iii)$, flare durations
are on average similar to the observations, even though with a
significantly smaller scatter, as would be expected with our simplifying
assumption of a single $\tau$ value. To correct this single
disagreement, in our last set of simulations (fourth row in
Fig.\,\ref{fig:flare_simul}) we allowed $\tau$ to span a
 range of
values, log-normally distributed with a mean=10\,ks and $\sigma$=1\,dex.
We are thus able to reproduce the observed distribution and range of
measured flare lengths, while the $VA$s and flare rates remain very
similar to those of our previous simulations set.

From the last two sets of simulations, the most successful at
reproducing the observed flare statistics, we conclude that flares are
likely {\em not} the dominant source of variability, as measured by the
$VA$s. This is particularly true at long timescales. In fact, in our
last simulation set  with a reasonable mix of flare decay-times that
well reproduce the measured flare durations, the run of $VA$s with
timescale remains rather flat, which is not compatible with the
observed sharp positive trends.

\section{Spectral Monte Carlo simulation}
\label{app:spMC}

Our preferred interpretation of the higher and more modulated X-ray
variability of CTTSs with respect to WTTSs is the intervention of
time-variable absorption of part of the X-ray emitting plasma. In
this scenario the observable X-ray spectra of CTTSs would be the sum of
two intrinsically similar components, undergoing very different
attenuations: while the emission of one plasma component would be
absorbed by interstellar material only, the emission from the other {\em
spatially distinct component would also intercept much thicker
circumstellar material.} This hypothesis seems to be contradicted by the
time-averaged X-ray spectra obtained by the COUP project, showing
neither signatures of such composed spectra, nor evidence of
significantly higher extinctions for CTTSs (\S\,\ref{sect:discPhyImp}).

To verify the significance of this lack of evidence, we have
investigated through Monte Carlo simulations whether the two above
signatures may or may not actually be detectable with data of the same
quality as those in our hands. More specifically, we simulated the
expected X-ray spectra in the two-absorptions scenario and verified
whether the composite nature of the resulting spectra could be detected
with the customary X-ray spectral analysis.

The simulated {\em Chandra} ACIS spectra were produced using XSPEC v12.7
\citep{arn96}, adopting an isothermal model (APEC) with kT=1.26\,keV,
quite typical of PMS stars, and assuming zero interstellar
extinction. The normalization was chosen so to obtain 1800 and 18000
counts (in two separate simulation sets). The lower value is somewhat
higher than the mean and the median number of counts obtained by COUP
for CTTS (as defined in this work): $\sim$450 and $\sim$1500,
respectively. The higher value is higher than the number of counts
collected for all CTTSs, but one.

For each simulation set, we simulated ``1T, two-$N_H$'' composed spectra
by varying $i)$ $f_{abs}$, the fraction of the coronal plasma that is
absorbed (from 0.1 to 1.0, in 0.1 steps), and $ii)$ $N_H$, the column
density of the absorbing material, (from $10^{21}$\,cm$^{-2}$ to
$3.2\times10^{23}$\,cm$^{-2}$, in 11 logarithmic steps). For each point
in this grid, we simulated 100 ACIS spectra, regrouped them so to have
SNR$\sim$5 per spectral bin, and fit them, within XSPEC, with an usual
two-temperature model with a single absorption (``2T, one-$N_H$''). We
judge the ability to detect the two absorptions from the goodness of
these latter fits, and specifically from the mean value of the null
probability,  (i.e. the probability of obtaining a higher $\chi^2$ value
given the model), $\overline{n.p.}$: high $\overline{n.p.}$ values
indicate that a model with a single $N_H$ is a good representation of
the composite spectrum and that the absorbed component cannot be
detected. To reject the single $N_H$ hypothesis, we would
require low values of $\overline{n.p.}$, e.g. $<$5\%  (a conservative
threshold).

\begin{figure}[!ht!]
\centering
\includegraphics[width=7.5cm]{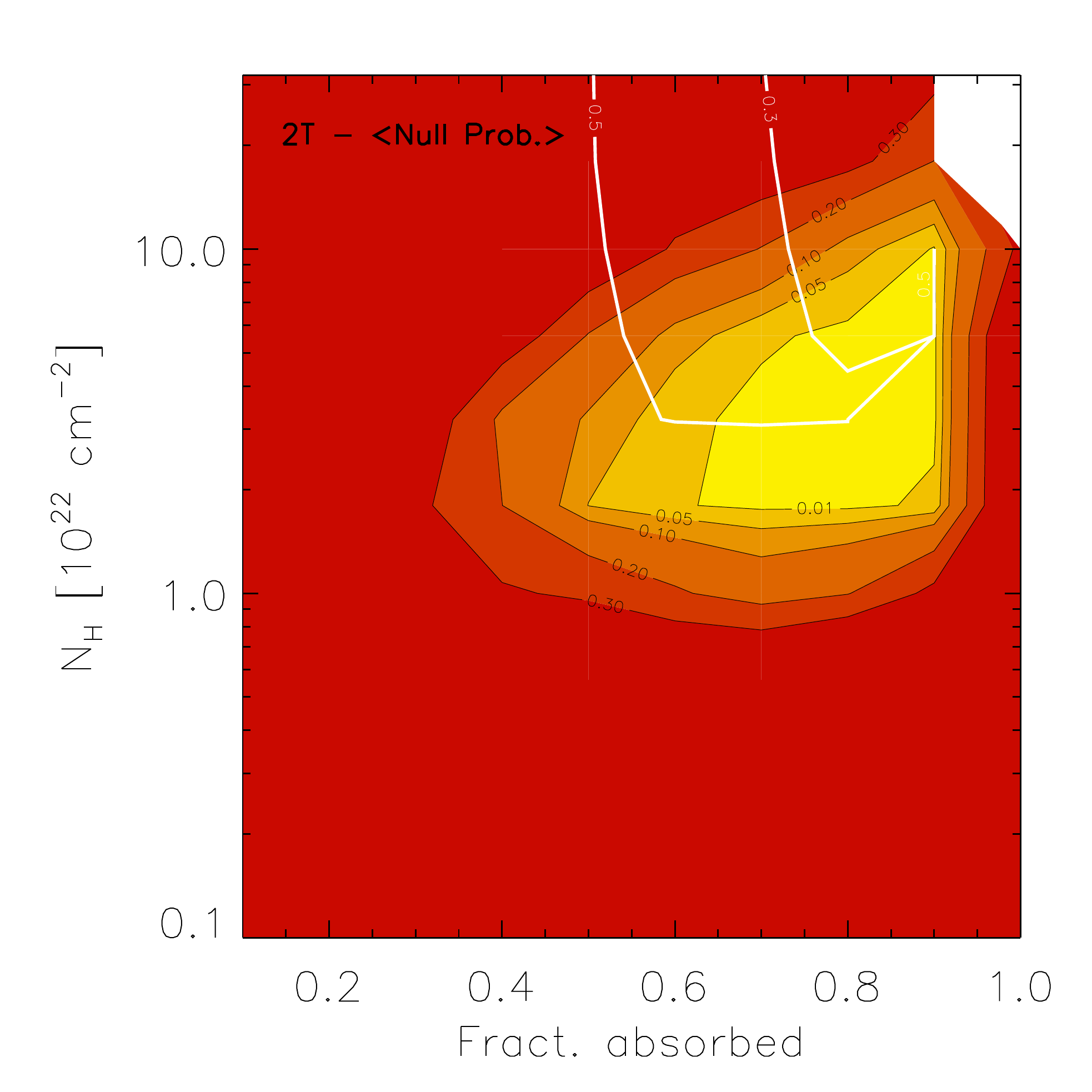}\vspace*{-0.3cm}
\includegraphics[width=7.5cm]{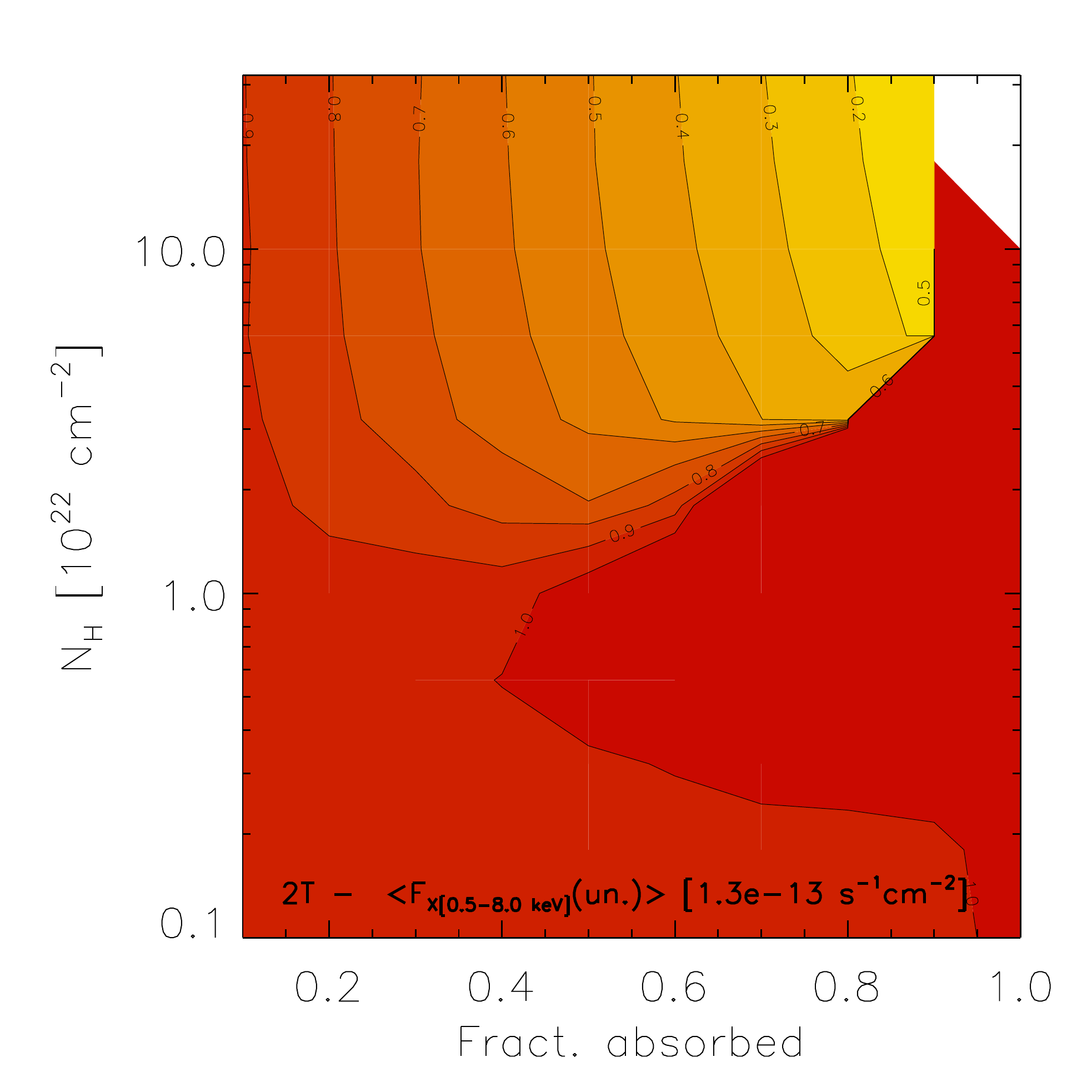}\vspace*{-0.3cm}
\includegraphics[width=7.5cm]{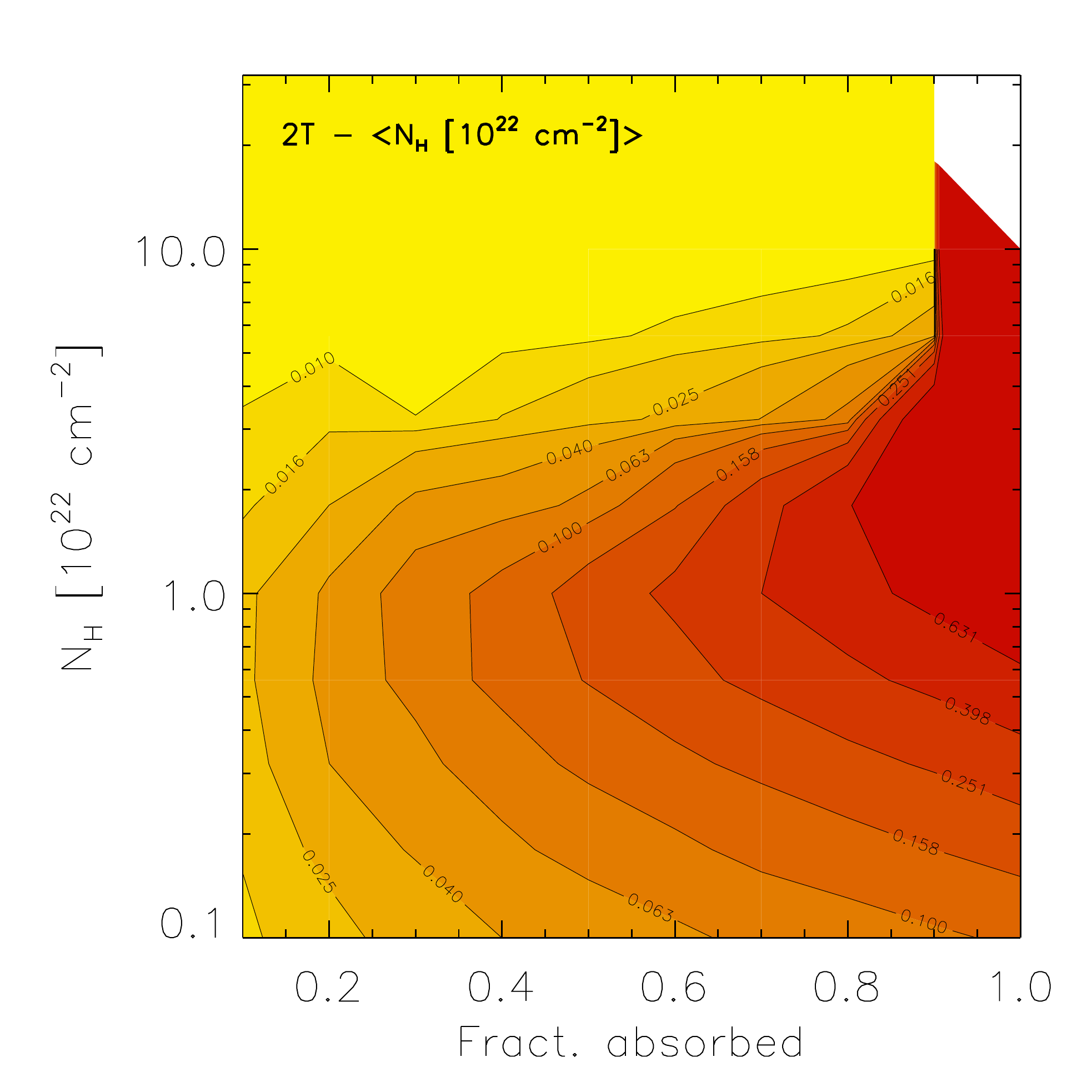} 
\caption{Results of the simulations of 1T, two-$N_H$ spectra
($\sim$18000 counts for $N_H=0$) fitted with 2T, one-$N_H$ models.
The black contours in the upper panel show the average null probability
as a function of the fraction of obscured emission measure, $f_{abs}$,
and the column density of the obscuring material, $N_H$. The middle and
lower panels show contour plots of the mean unabsorbed
flux (in the 0.5-8.0\,kV band, in units of 1.3$\times 10
^{-13}$\,s$^{-1}$cm$^{-2}$), and of the mean $N_H$ (units $10
^{22}$\,cm$^{-2}$), both obtained from the spectral fits. The two white
contours in the upper panel are repeated from the middle panel and show
the loci corresponding to a 50\% and 70\% reduction in measured flux.}
\label{fig:2nh_simul}
\end{figure}

Figure\,\ref{fig:2nh_simul} is a contour plot showing $\overline{n.p.}$
derived from the simulations with the brighter template spectrum (18000
counts for the unabsorbed case) in the $f_{abs}$-$N_H$ plane. We see
that for much of the plane $\overline{n.p.}$ is larger than 5\%,
indicating that even with high statistics we are unable to detect the
existence of two absorption components for most $f_{abs}$-$N_H$
combinations. The same plot for the case of the fainter template
spectrum (1800 counts) shows that $\overline{n.p.}>$10\% for all
$f_{abs}$-$N_H$ pairs.  The middle panel in Fig.\,\ref{fig:2nh_simul} 
shows the mean unabsorbed flux obtained from the fits with 2T,
one-$N_H$ spectral models, at each position in the grid and relative
to the flux for the $N_H$=0 case. To obtain a reduction of the
flux by a factor of 2-3, i.e. on the order of the mean $L_X$ difference
between WTTS and CTTSs, $N_H$ must be greater than $\sim
3\times$10$^{22}$\,cm$^{-2}$ and $f_{abs}$=0.5-0.7. The contour lines
relative to a decrease in flux by a factor of 2 and 3 are also reported
in white in the  $\overline{n.p.}$ contour plot (upper panel). Note
that there is ample room in the $f_{abs}$-$N_H$ plane for having a
decrease in retrieved flux on this order, and at the same time being
unable to detect the second absorption. The bottom panel of
Fig.\,\ref{fig:2nh_simul} shows the mean $N_H$ obtained from the
spectral fits, showing that in the same region of the parameter space
discussed above, not only is it impossible to detect the second $N_H$,
but the value of the best fit $N_H$ obtained from a single-$N_H$ model
is very low, basically that of the unabsorbed component. We also note
that in the same region of the plane, the values of the two
temperatures obtained from the spectral fits with 2T, one-$N_H$
models, are in line with observed values (not shown).

We did not try to fit the spectra with more than two thermal components
or to allow plasma abundances to vary in the spectral fits. Therefore
the combinations of $f_{abs}$-$N_H$  values for which a single-$N_H$
model is not adequate to fit the two-$N_H$ spectra may be even more
limited than found above. Moreover, we neglected systematic
uncertainties, e.g. on the models and on the calibration of the spectral
responses, which, if considered, might make the rejection of a one-$N_H$
model even harder.

We conclude that the scenario in which a large part of the coronal
emission is obscured by stable or time-varying gaseous structures is
compatible with the observed spectra, showing neither evidence of double
absorptions nor substantially higher best fit $N_H$ values. Moreover,
assuming that WTTS and CTTS have similar coronae and that the difference
in observed flux is due to partial obscuration of the corona, we must
require that the obscuring material has $N_H$ greater than a few
10$^{22}$\,cm$^{-2}$.

\end{document}